\documentclass[]{interact}

\usepackage{epstopdf}
\usepackage[caption=false]{subfig}

\usepackage[numbers,sort&compress]{natbib}
\bibpunct[, ]{[}{]}{,}{n}{,}{,}
\makeatletter
\def\NAT@def@citea{\def\@citea{\NAT@separator}}
\makeatother
\usepackage[colorlinks=true,linkcolor=blue,citecolor=blue]{hyperref}
\theoremstyle{plain}

\theoremstyle{definition}

\theoremstyle{remark}

\begin{document}

\articletype{REVIEWS}

\title{The Rise of Compositionally Complex Perovskite Oxides in Quantum Materials Research}

\title{Extreme Disorder in Crystalline Perovskite Oxide: a New Paradigm in Quantum Materials Research}

\author{Srimanta Middey, Nandana Bhattacharya, Rukma Nevgi, Suresh Chandra Joshi, and Subha Dey}
\author{
\name{Srimanta Middey\textsuperscript{a}\thanks{CONTACT Srimanta Middey: Email: smiddey@iisc.ac.in}, Nandana Bhattacharya\textsuperscript{a}, Rukma Nevgi\textsuperscript{a}, Suresh Chandra Joshi\textsuperscript{a}, and Subha Dey\textsuperscript{a}}
\affil{\textsuperscript{a}Department of Physics, Indian Institute of Science, Bengaluru 560012, India}
}

\maketitle

\begin{abstract}

Perovskite oxides ($AB$O$_3$) have long been central to the advancement of modern condensed matter physics, owing to their rich and tunable electronic and magnetic properties. The quest to understand their various entangled phases has spurred the development of both cutting-edge experimental tools and innovative theoretical frameworks. In recent times, the emergence of high entropy oxides - materials in which five or more elements share a single crystallographic site - has introduced a powerful new paradigm in materials design. Embedding such extreme chemical disorder within the perovskite framework has opened vast opportunities for realizing novel physical phenomena inaccessible in conventional oxides. This review surveys the rapid advances in the synthesis, characterization, and exploration of the electronic and magnetic properties of compositionally complex perovskite oxides, offering key insights and highlighting promising avenues for future research.

\end{abstract}


\section{Introduction}
The advancement of civilization is closely tied to the discovery and development of new material platforms—such as the Copper Age, Bronze Age, Iron Age, and, in modern times, the Silicon Age. We are on the brink of a new era - quantum technology,  which holds the potential to revolutionize fields such as computing, cryptography, communication, and sensing by enabling unprecedented processing power and security. This advancement hinges on the creation and engineering of quantum materials (abbreviated as Q-materials in this article) with the desired functionality. Traditional Q-materials design principles generally consider disorder as something to be avoided to realize intriguing phenomena.
For example, following the discovery of the integer quantum Hall effect~\cite{Klitzing:1980p494}, the enhancement of the electron mobility in GaAs-AlGaAs heterojunctions led to the discovery of the fractional quantum Hall effect~\cite{Tsui:1982p1559}.
The continuous developments in thin film growth techniques have enabled the realization of two-dimensional electron systems with ultra-high mobility hosting new fractional Hall states, which are promising platforms for the topological quantum computing~\cite{Chung:2021p632,Wang:2025p046502}.
Electron hydrodynamic flow, originally hypothesized by Gurzhi in 1963~\cite{Gurzhi:1963p045443}, provides another example of emergent behavior upon disorder reduction, and has only recently been observed thanks to advances in ultra-clean 2D material fabrication and measurement techniques~\cite{Moll:2016p1061,Vool:2021p1216,Majumdar:2025universality}. Similarly, material improvements are now recognized as potent methods for enhancing superconducting qubit lifetimes and coherence times~\cite{Ganjam:2024p3687}. 
  Beyond its often-detrimental effects, disorder can play a pivotal role in realizing a range of intriguing phenomena such as Anderson localization~\cite{Anderson:1958p1492}, glassy phase (structural glass, spin glass, electron glass)~\cite{Amir:2011p235,Mott:2012,Binder:1986p801}, many-body localization~\cite{Abanin:2019p021001}, etc. Furthermore, the interplay between disorder and topology has recently gained attention, as moderate disorder can induce topologically nontrivial phases~\cite{Liu:2025p108}. Thus, the disorder can also be strategically employed as a control parameter to engineer collective phenomena.

\begin{figure}
    \centering
	\includegraphics[width=.82\textwidth] {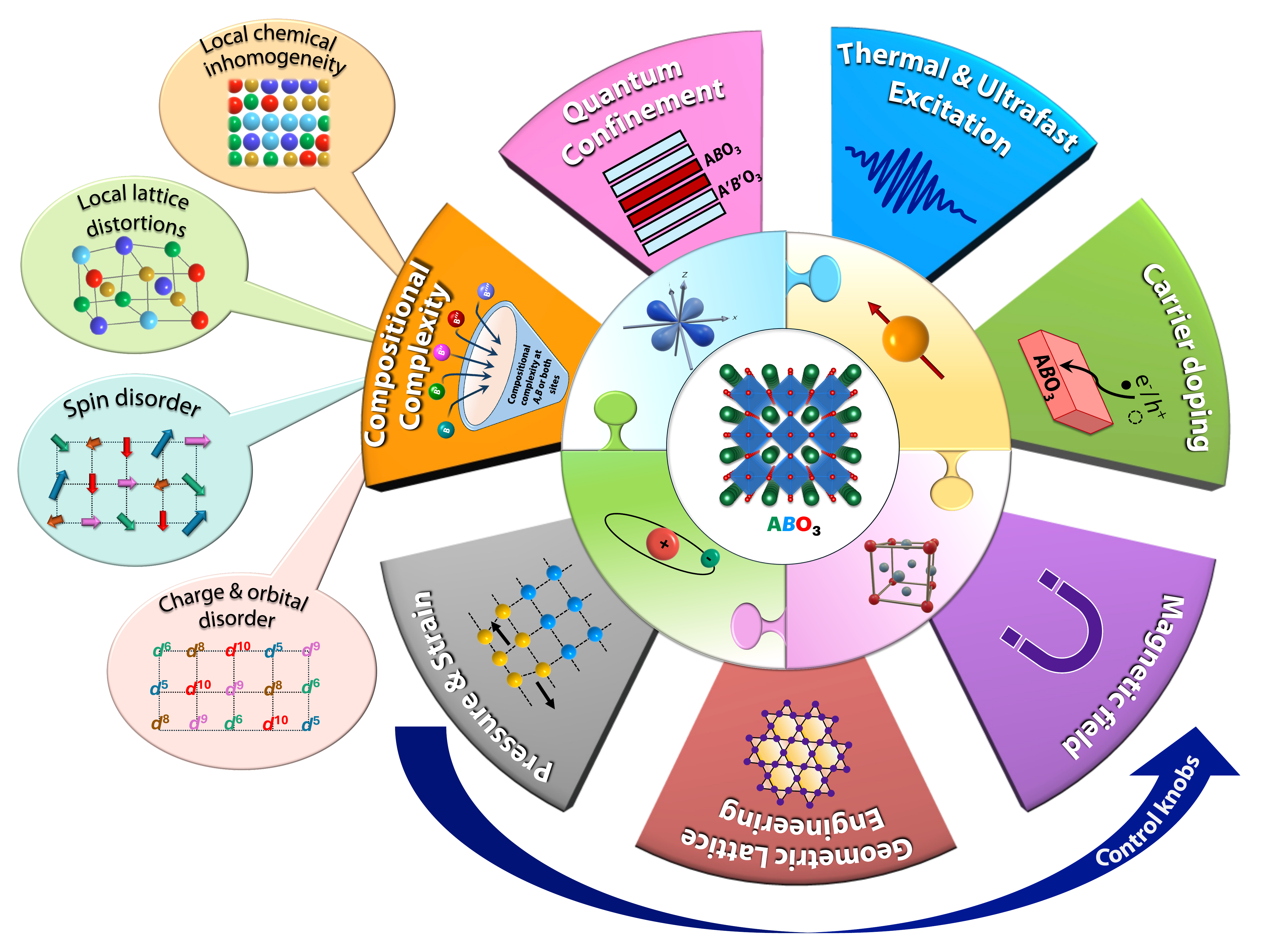}
	\caption{\label{Fig1}
    Schematic illustration to highlight the tuning parameters for perovskite oxide properties such as carrier doping, strain engineering, quantum confinement, thermal excitation, etc. `Compositional complexity' is highlighted as an additional, recently recognized control knob that encapsulates the capacities of introducing chemical inhomogeneity, lattice distortions, spin, charge, and orbital disorders at the local scale.}
\end{figure}

The highly tunable disorder inherent in compositionally complex materials (CCMs) has positioned them as a significant platform for investigating disorder-driven phenomena in recent years. A number of review articles are already available on this broad topic~\cite{Sen:2024p3694,Oses:2020p2058,Kotsonis:2023p5587,Wang:2023p2201138,Betti:2025p2401199,Liu:2022pe12261,Bao:2024p23179,Ma:2021p2883,Sarkar:2019p1806236,Aamlid:2023p5991,Mazza:2024p230501,Musico:2020p040912,Brahlek:2022p110902,Schweidler:2024p266,Zou:2024p34492}. This article specifically focuses on recent advancements in perovskite oxide-based CCMs.  The chemical formula for perovskite oxides is $AB$O$_{3}$, where the $A$-site can be alkali metal ions, alkaline earth metal ions, or rare-earth ions, and the $B$-site is occupied by transition metal ions.
Perovskite oxides are among the earliest recognized Q-materials, displaying a rich spectrum of electronic and magnetic phenomena, including metal-insulator transitions (MIT), superconductivity, quantum magnetism, orbital ordering, charge ordering, multiferroicity, etc~\cite{Goodenough:1963magnetism,khomskii:2014,Imada:1998p1039,Cooper:2003book,Tokura:2017p1056,Ahn:2021p1462}. For over 75 years, perovskite oxides have stood as a cornerstone in modern condensed matter physics, not only for their diverse functional properties but also for their exceptional tunability under external stimuli including electric and magnetic fields, pressure, and light pulses, etc [Fig. \ref{Fig1}]~\cite{Cox:2010book,khomskii:2014,Imada:1998p1039,Tokura:2000p462,Ramirez:1997p8171,Tokura:2006p797,Disa:2021p1087,Mao:2018p015007}. Manipulating both cationic sublattices along with the anionic sublattice have expanded the phase diagram of perovskite oxides into more exotic phases. Consequently, carrier doping via chemical substitution at both $A$ and $B$ sites, oxygen vacancy creation, electrostatic gating, and ionic liquid gating represent another crucial frontier for control~\cite{Ahn:2006p1185,Goldman:2014p45}. The underlying core of these versatile tunable properties is the strong intercoupling among lattice, charge, orbital, and spin degrees of freedom. The tremendous advancements in thin film growth technologies over the last two decades have unleashed a suite of powerful new control methodologies — epitaxial strain, designer heterointerface, quantum confinement, geometrical lattice engineering, etc.— which have, in turn, unlocked a wealth of emergent phenomena.  A detailed discussion of these effects is beyond the scope here, as several insightful review articles are already available on the subject~\cite{Hwang:2012p103, Chakhalian:2014p1189,Schlom:2014p118, Middey:2016p305, Liu:2016p133, Schlom:2008p2429, Zubko:2011p141, Chakhalian:2012p92, Ramesh:2019p257, Fernandez:2022p2108841}. Beyond these control parameters [Fig. \ref{Fig1}], increasing the compositional complexity establishes a new paradigm within the $AB$O$_3$ series, which forms the central focus of this review.

Rather than attempting an exhaustive survey, we present a selection of results to illustrate the concepts, methodologies, and physical principles relevant to the impact of compositional complexity within. 
We apologize for any unintentional oversights. This paper is structured as follows: we begin with a discussion about CCMs. We then concentrate on the electronic and magnetic properties of compositionally complex perovskite oxides, exploring their relationship with structural distortions. We have intentionally excluded extensive discussions about applications of these materials for practical uses, as this topic is well-covered in several existing reviews~\cite{Sarkar:2019p1806236,Schweidler:2024p266,Wang:2023p2201138,Betti:2025p2401199,Liu:2022pe12261}. Finally, we provide an outlook, highlighting some promising new directions.

\section{What are compositionally complex materials?}
The field of CCMs finds its roots in the pioneering 2004 work on high-entropy alloys (HEAs) by Yeh and Cantor \textit{et al.}, which demonstrated the viability of a single-phase material containing five or more constituent elements~\cite{Yeh:2004p299, Cantor:2004p213}. For completeness, we briefly outline the design principle of HEAs, referring to Ref.~\cite{Murty:2019y2019} for a comprehensive treatment. A critical determinant in alloy formation is the total mixing entropy, which encompasses contributions from configurational, vibrational, spin, and electronic entropy. As the configurational entropy is the predominant among these~\cite{Swalin:1962p2308, Fultz:2010p247}, it is used as a hallmark to demarcate various alloy categories based on their elemental configurations.

\begin{figure}[!b]
    \centering
	\includegraphics[width=1.0\textwidth]{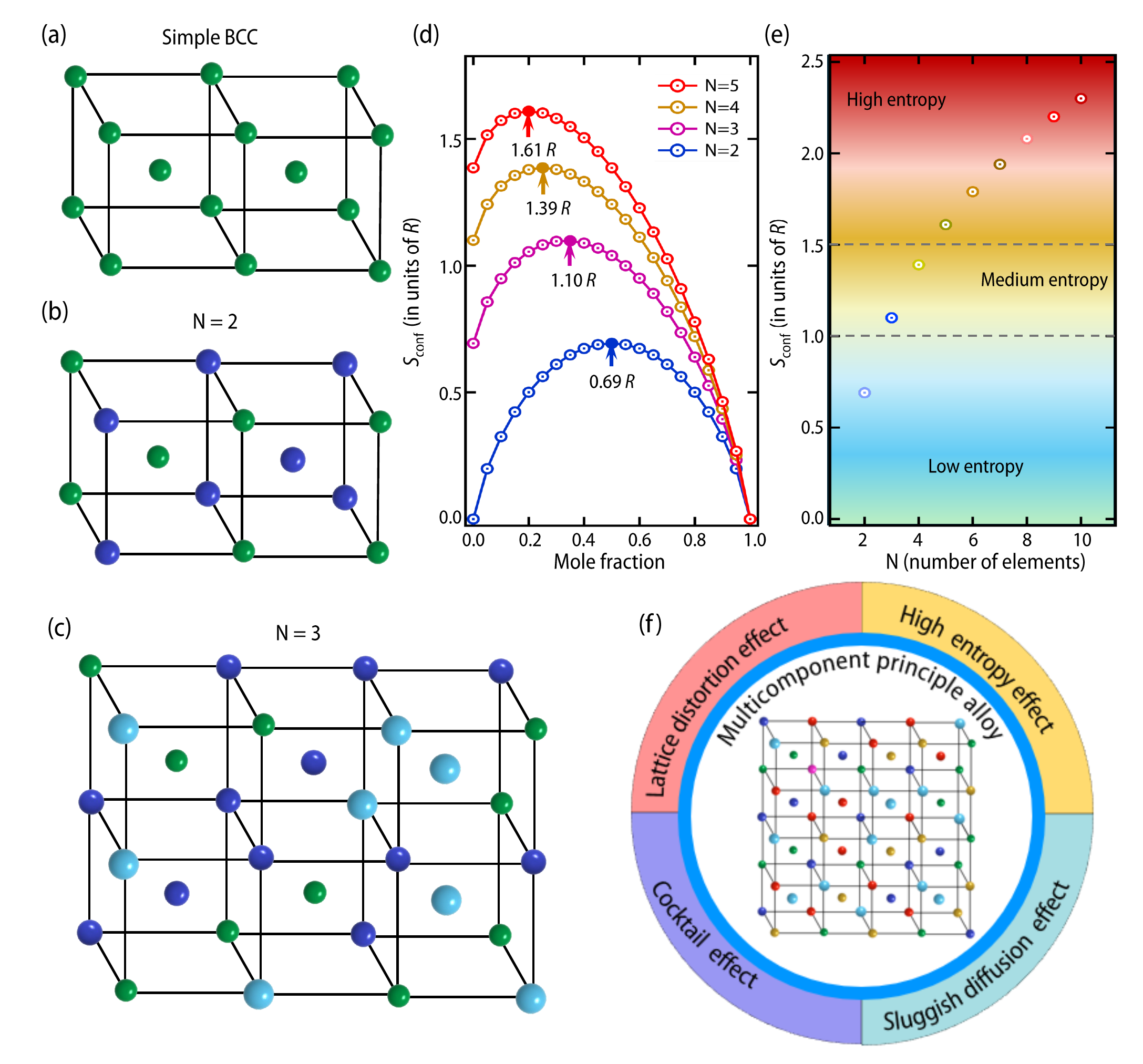}
	\caption{\label{Fig2} Concept of high entropy alloy. Body-centered cubic (BCC) structure schematics for increasing number ($N$) of elements (a) $N$ = 1,  (b) $N$ = 2, and (c) $N$ = 3.  (d) Configurational entropy change, $S_\mathrm{conf}$ = $-R (xlnx+(1-x)ln(\frac{1-x}{N-1}))$, calculated by varying the concentration $x$ of one species while keeping the remaining $N-1$ components equimolar, shown for $N$ = 2, 3, 4, 5. 
    (e) Maximum configurational entropy as a function of the number of elements ($N$= 2 - 10)~\cite{Aamlid:2023p5991}. (f) The four core effects of high-entropy design in multi-component alloys: (i) the high-entropy effect, which stabilizes single-phase solid solutions and suppresses brittle intermetallic formation \cite{Yeh:2013p1759}; (ii) severe lattice distortion, which enhances hardness and strength while reducing thermal effect; (iii) sluggish diffusion, leading to slower grain growth and finer precipitates that improve strength and rigidity \cite{Liu:2013p526} and (iv) the cocktail effect, which introduces synergistic atomic-scale to microscale interactions that benefit high-temperature performance \cite{Ranganathan:2003, Murty:2019y2019}. Collectively, these effects substantially enhance the mechanical and thermal properties and plasticity of HEAs \cite{George:2019p515, Hsu:2024p471, Zhang:2013p1455, Chen:2021p4953, Li:2016p227}.
    }
\end{figure}

From statistical thermodynamics, the configurational entropy change per mole, $\Delta S_{\text{conf}}$, during the formation of a solid solution from $n$ elements is given by~\cite{Aamlid:2023p5991},
 
\begin{equation}
\Delta S_{\text{conf}} =k_B\ \mathrm{ln}(\Omega)= R\ \mathrm{ln}\left[\frac{N!}{\prod_in_i!}\right]  \approx -R \sum_{i=1}^{n} x_i \ln x_i
\end{equation}
where $x_i$ (=$n_i$/$N$) is the mole fraction of the component $i$, $N$=$\sum_i n_i$, $R$ is the universal gas constant 8.314 J/mol-K.
It can be further shown that the $\Delta S_\mathrm{conf}$ is maximum [=$R$ ln $N$] when the mole fraction of the individual elements are equiatomic as depicted in Fig.~\ref{Fig2}(a)-(d).  Based on the maximum attained value of $\Delta$$S_\mathrm{conf}$, alloys have been classified into low, medium and high entropy alloys [Fig.~\ref{Fig2}(e)] where the $\Delta$$S_\mathrm{conf}$ are $\leq$ 1 $R$, between 1 and 1.5 $R$ and $\geq$ 1.5 $R$, respectively. As evident from Fig.~\ref{Fig2}(d), the $N$ = 5 alloy exhibits a maximum $\Delta$$S_\mathrm{conf}$ $\sim$ 1.61 $R$ $>$ 1.5 $R$, thus setting the lower bound to the HEA systems.  
The thermodynamic favorability of high-entropy phases at high temperatures stems from the dominance of the $-T \Delta S$ term in the Gibbs free energy ($\Delta G = \Delta H - T \Delta S$). 
Following the pioneering synthesis of CrMnFeCoNi~\cite{Yeh:2004p299,Cantor:2004p213}, HEAs have rapidly become a vital field in metallurgy, largely due to their numerous advantages over conventional alloys [Fig.~\ref{Fig2}(f)]~\cite{Murty:2019y2019,Hsu:2024p471} and their tendency to form simple phase structures (bcc, hexagonal close packed (hcp), and face-centered cubic (fcc))~\cite{Sun:2019p090301}. To date, over 90 different types of HEAs exhibiting these structures have been reported~\cite{Ye:2016p349}.

\begin{figure}[!b]
    \centering
	\includegraphics[width=1.0\textwidth] {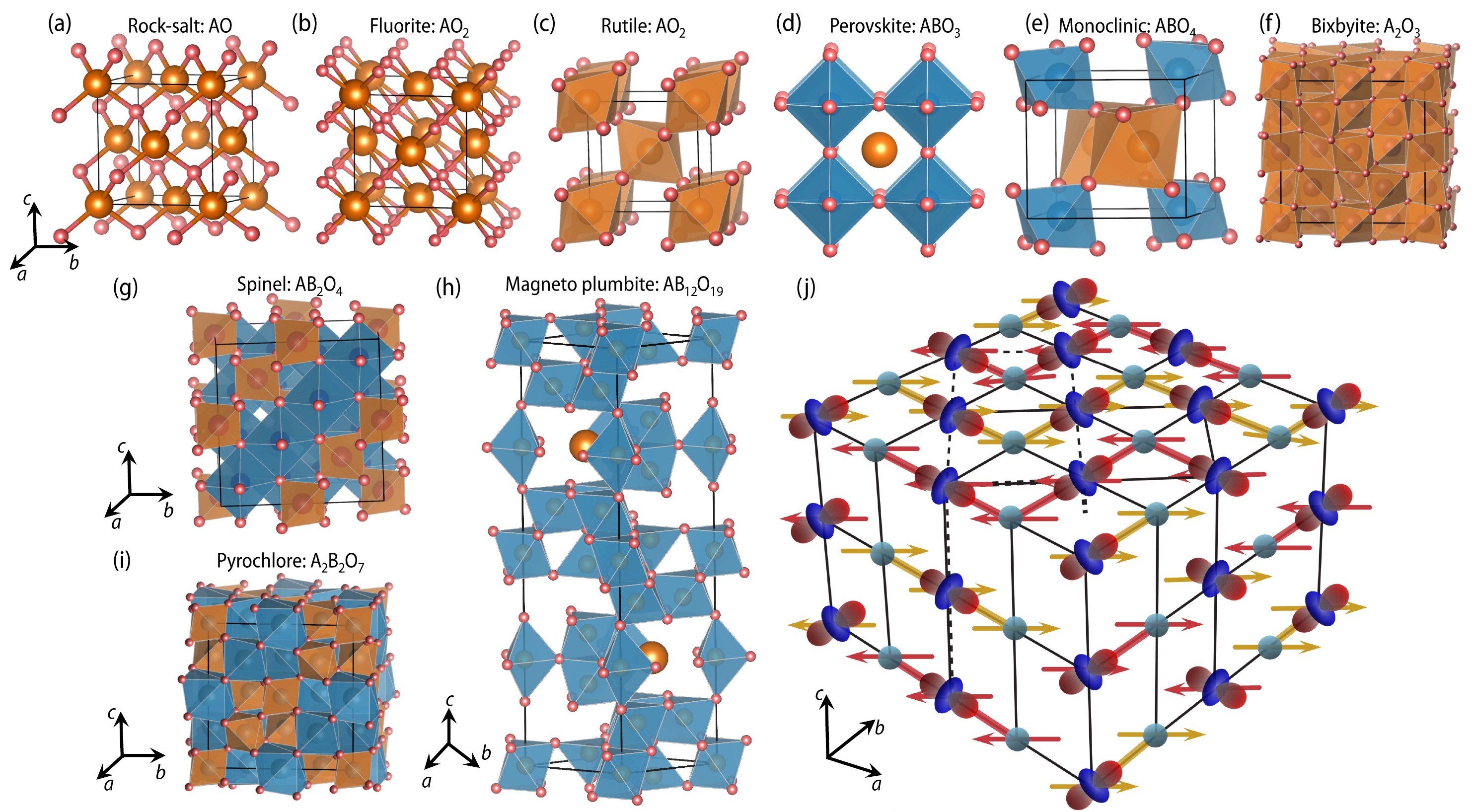}
	\caption{\label{Fig3} (a) Rock-salt, (b) fluorite, (c) rutile, (d) perovskite, (e) monoclinic, (f) bixbyite, (g) spinel, (h) pyrochlore, and (i) magnetoplumbite structures commonly adopted by TMOs. 
(j) Schematic of a half-doped perovskite manganite illustrating the coupled charge, orbital, and CE-type antiferromagnetic ordering, adapted from Ref.~\cite{Ulbrich:2011p094453}.}
\end{figure}

The high-entropy concept evolved to explore the impact of compositional complexity on a multitude of functional properties in ceramics such as oxides, carbides, borides, nitrides, and sulfides~\cite{Akrami:2021p100644}. High-entropy ceramics differ from alloys by having separate cation and anion sublattices. (Mg,Co,Ni,Cu,Zn)O was the first high entropy oxide (HEO), reported in 2015 by the pioneering work of Rost \emph{et al}~\cite{Rost:2015p8485}.  This single-phase compound was obtained by heating a mixture of its constituent binary oxides above 1150 K and subsequently quenching, a notable achievement given that these binary oxides do not typically form such a solid solution. Following this initial demonstration of an entropy-stabilized oxide with a rocksalt structure, the field has expanded significantly over the past decade, with reports of HEOs exhibiting a variety of other crystal structures [Fig.~\ref{Fig3}], including fluorite, bixbyite, perovskite, spinel, magnetoplumbite, pyrochlore, rutile, and defect fluorite, etc.~\cite{Musico:2020p040912}. As some of these crystallographic structures have multiple cationic sublattice, the overall $\Delta S_\mathrm{conf}$ can be estimated  by generalizing equation (1) to incorporate a
second summation across sublattices~\cite{Sen:2024p3694}:
\begin{equation}
  \Delta S_\mathrm{conf}=-R\sum_{i=1}^n m_j\sum_{i=1}^n x_{i,j}\ \mathrm{ln}\ x_i
\end{equation}
 Here $m_j$ is the multiplicity of sublattice $j$ and $x_{i,j}$
denotes the proportion of element $i$ present on sublattice $j$.

 Although a universally accepted definition for HEOs remains elusive, a prevalent criterion for their classification involves the presence of five or more cations in near-equimolar proportions within a single-phase crystal lattice exhibiting well-defined structural symmetries~\cite{Miracle:2017p448, Sarkar:2020p43, Aamlid:2023p5991}. Terms such as HEO, entropy stabilized oxides, and multicomponent oxide are frequently used interchangeably in the literature, despite having subtle distinctions. For instance, not all HEOs are necessarily entropy-stabilized, and examples of entropy-stabilized low-entropy oxides also exist~\cite{Aamlid:2023p5991}. In this article, we adopt the broader classification of `compositionally complex oxides' (CCOs), under which HEOs are categorized as a subclass~\cite{Brahlek:2022p110902}. 
Our focus here is on CCOs that crystallize in the perovskite structure, which we refer to as `compositionally complex perovskite oxides' (CCPOs) throughout this paper.

\section{Perovskite oxide}
The term `perovskite' is named in honor of the distinguished Russian mineralogist Count Lev Alexevich von Perovski [See ~\cite{George:2020} for a detailed account on perovskites]. The first mineral identified with this structure was CaTiO$_3$, discovered in the Urals in 1839 by Gustav Rose, subsequently other naturally occurring perovskite oxides include FeTiO$_3$, MgSiO$_3$, etc. Historically, BaTiO$_3$ was the first synthetic perovskite, developed during World War II, and it continues to be widely explored for its ferroelectric and piezoelectric functionalities. On the other hand, SrTiO$_3$, the foundational cornerstone for modern oxide electronics, was the first recognized perovskite band insulator which also happened to be the first oxide superconductor~\cite{Collignon:2019p25,Schooley:1964p474}.

In $AB$O$_3$ compounds, the $B$ cation largely dictates the material's electronic and magnetic properties, whereas the $A$ cation sublattice is principally responsible for stabilizing the structure and determining the valency of the $B$ cation. Theoretical predictions dictate that there are 49 $A$-site and 68 $B$-site elements on the periodic table that are capable of forming perovskites, making it $\sim$ 3000 possible combinations. Now, introducing compositional complexity at the $A$ site and/or $B$ site as a control parameter multiplies these combinations by several folds, introducing richer paradigms (Fig.~\ref{Fig1}). The profound high disorder  breaks the symmetries of the spin, charge, and orbital sectors across all length scales, simultaneously introducing  local disorder distortions remarkably preserving an overall uniform crystal structure on larger scales~\cite{Mazza:2024p230501}. Inevitably, this dramatic impact is fueling intense and rapid efforts to study these CCPOs since the first report by Jiang \emph{et al.}~\cite{Jiang:2018p116}, both in bulk and epitaxial thin film forms, which we have summarized in this article.

\subsection{Structural distortions in perovskites} For the ideal cubic perovskite, the relation between the lattice parameter $a$ and the ionic radii of the cations ($r_{A}$, $r_{B}$) and the oxygen ($r_{O}$) is  $a$ =$\sqrt{2}$($r_{A}$+$r_{O}$)=2($r_{B}$+$r_{O}$)~\cite{George:2020}
But, for most of the perovskite compounds this relation does not hold, and the oxygen octahedra tend to distort, leading to a less symmetrical structure. The assessment of the tendency to distort can be understood by Goldschmidt's tolerance factor ($t$)~\cite{Goldschmidt:1926p477}, which is defined as: $t=\frac{(r_{A}+r_{O})}{\sqrt{2}(r_{B}+r_{O})}$. For the ideal cubic system, $t$=1, yet the cubic perovskite structure remains stable for 0.9 $<t<$ 1 (certain compounds relax to a rhombohedral structure). The perovskites have a preference for rhombohedral or orthorhombic structures, two of the most common perovskite structures when $t$ is less than 0.9, and a hexagonal structure when $t$ is greater than 1. Some of the common types of perovskite structure and their associated space groups have been summarized in Fig.~\ref{Fig4}(a). Moreover, $A$-site displacements, octahedral tilts, and octahedral rotations are very common in $AB$O$_3$ oxides, and these distortions may be expressed using a convenient shorthand notation developed by A. M. Glazer~\cite{Glazer:1975p756,Glazer:1972p3384}. In this convention, the ideal cubic perovskite has  the Glazer notation of $a^0a^0a^0$. The common orthorhombic structure has an $a^-b^+c^-$ rotational pattern in pseudocubic notation. Another common symmetry, the rhombohedral structure, has an $a^-a^-a^-$ tilt pattern indicating out-of-phase rotation with the same magnitudes along all three directions. It is important to note that, by reducing the symmetry, the transition between many different perovskite structures is feasible in a seamless manner. However, this is not allowed for any arbitrary combination, and Fig.~\ref{Fig4}(a) provides a summary of all the various space groups that are associated with $AB$O$_3$ oxides that undergo pure octahedral rotations, as well as the subgroup relations that exist between those space groups~\cite{Howard:1998p782}.

In addition to tilt and rotation-driven structural changes, the perovskite \textit{B}O$_6$ octahedra may also exhibit other intrinsic  distortions.  One very common distortion is the Jahn–Teller effect, observed in systems with electronically degenerate $d$ orbitals such as Mn$^{3+}$ in LaMnO$_3$, which drives an elongation or compression of the octahedra to lift the degeneracy and stabilize the electronic configuration~\cite{Tokura:2000p462}. Another well-known distortion is the breathing mode, observed in systems like $RE$NiO$_3$, wherein the alternate octahedra expand and contract, giving rise to inequivalent $B$–O bond lengths~\cite{Middey:2016p305}. A third variant is antipolar or polar off-centering distortions, where the $B$-site cation shifts from the octahedral center (such as Ti$^{4+}$ displacement in BaTiO$_3$), producing local dipoles that couple to ferroic order parameters~\cite{Rabe:2007p1}.

Therefore, an accurate determination of the nature and extent of octahedral distortions is crucial for understanding structure–property relationships in complex oxides. This has been accomplished through a combination of diffraction, microscopic, and spectroscopic techniques, each offering complementary structural information across different length scales. X-ray diffraction (XRD) provides insight into the long-range crystallographic symmetry and allows identification of the octahedral tilt system via refined space-group analysis~\cite{Rietveld:1969p65}. To access shorter correlation lengths, pair distribution function (PDF) analysis probes intermediate length scale ($\sim$10-100 \AA), capturing deviations from the average perovskite structure and revealing local distortions that may be invisible to conventional diffraction~\cite{Billinge:2008p1695}. At the atomic scale, scanning transmission electron microscopy (STEM) directly visualizes octahedral rotations, cation displacements, and antisite disorder, enabling spatial mapping of structural heterogeneity and symmetry breaking in nanoscale~\cite{Muller:2009p263,Pennycook:2012p28}. Complementary to these, extended X-ray absorption fine structure (EXAFS) spectroscopy provides element-specific quantitative information on local bond lengths, coordination geometry, and Debye–Waller factors, offering a direct measure of lattice relaxation and octahedral deformation around specific cations~\cite{Rehr:2000p621,Nevgi:2025p7038}. Together, these multiscale probes enable a comprehensive understanding of structural distortions in oxides. A detailed discussion on local structural modulations and their implications for functionality in HEOs can be found in the recent review by Barber \textit{et al.}~\cite{Barber:2025}. Here, we will discuss briefly on some of the local structural studies w.r.t. CCPOs in the following section to establish an introductory ground for their structure-property relations.

\section{How does compositional complexity affect global and local structure in CCPOs?}

Since the bond lengths and bond angles in a simple $AB$O$_3$ perovskite structure are primarily governed by the ionic radii of the $A$ and $B$-site cations, introducing compositional complexity at either of the sites or both inevitably induces local strain. This raises a fundamental question: how do the $B$–O bond lengths and $B$–O–$B$ bond angles respond at the local scale to such compositional disorder? 
In this context, atomic-scale chemical fluctuations emerge as an important consideration, as demonstrated by Su \textit{et al.}~\cite{Su:2022p2358} in bulk samples. Their annular bright-field STEM (ABF-STEM) analysis revealed a broad distribution of $B$-O-$B$ bond angles, with the extent of this distribution being more pronounced in samples exhibiting lower tolerance factor. Since such chemical fluctuations have been observed in systems with compositional complexity at the $B$-site [La(Cr$_{0.2}$Mn$_{0.2}$Fe$_{0.2}$Co$_{0.2}$Ni$_{0.2}$)O$_3$, Gd(Cr$_{0.2}$Mn$_{0.2}$Fe$_{0.2}$Co$_{0.2}$Ni$_{0.2}$)O$_3$],  or both [(La$_{0.2}$Nd$_{0.2}$Sm$_{0.2}$Gd$_{0.2}$Y$_{0.2}$)(Cr$_{0.2}$Mn$_{0.2}$Fe$_{0.2}$Co$_{0.2}$Ni$_{0.2}$)O$_3$], it is highly plausible that bond-angle variability and local strain heterogeneity are intrinsic features of CCPOs.

\begin{figure}
    \centering
	\includegraphics[width=1.0\textwidth] {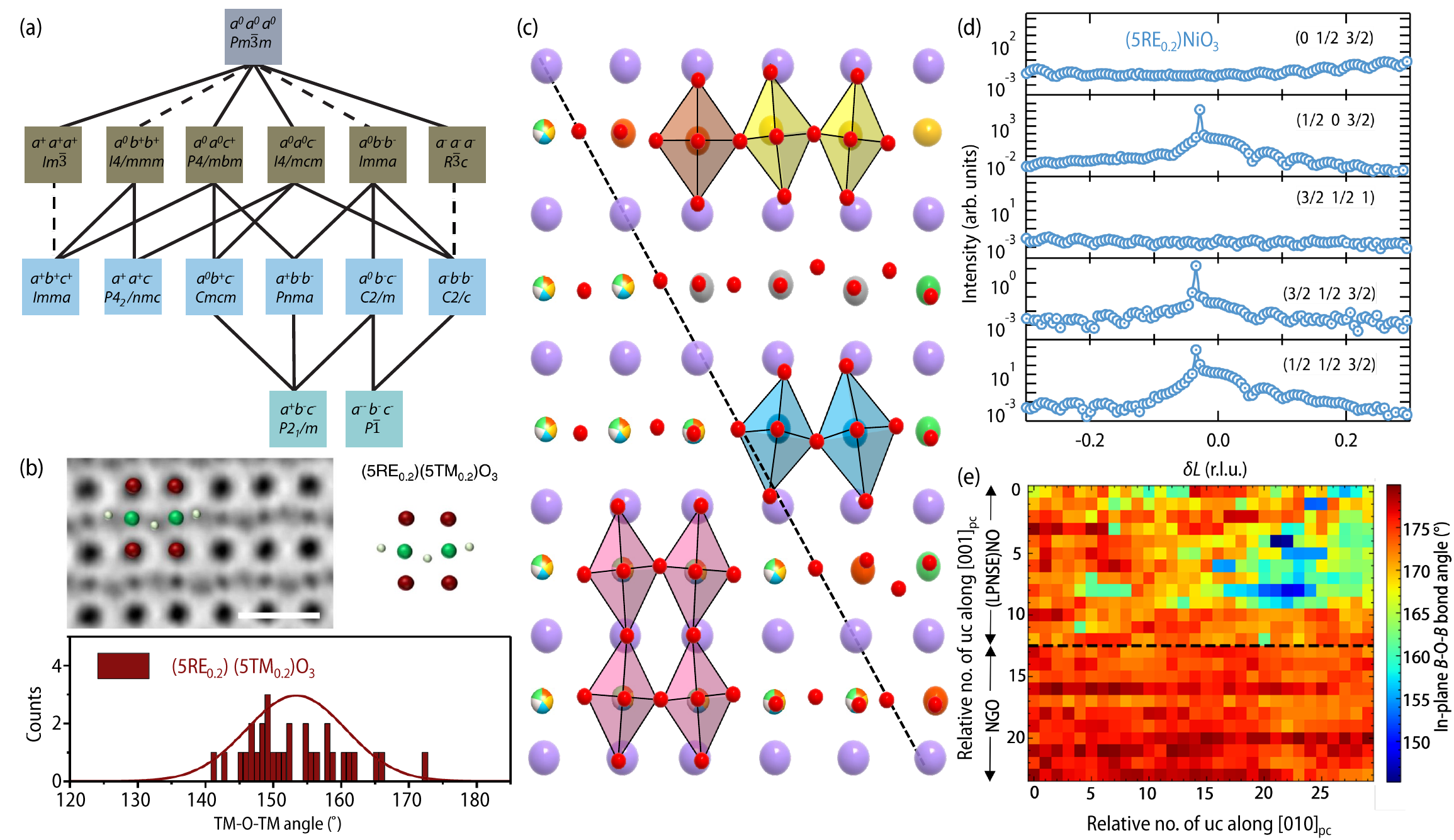}
	\caption{\label{Fig4} Effect of compositional complexity on local structure in perovskite oxides: (a) Space groups corresponding to different octahedral rotational patterns in perovskites, along with their Glazer notations. The continuous lines denote first-order transitions, while dashed lines indicate second-order–like transitions. The absence of connecting lines between two symmetries signifies the lack of a group–subgroup relationship (Adapted from Ref.~\cite{Howard:1998p782} with permission). (b) Atomic structure and compositional fluctuation revealed by annular bright-field (ABF) imaging along the [110] zone axis (upper panel), and the distribution of TM–O–TM bond angles in (5RE$_{0.2}$)(5TM$_{0.2}$)O$_3$ (lower panel) (c) Compositional fluctuation with nanoscale ordering and associated local octahedral distortions in an orthorhombic CCPO. Fig.~\ref{Fig4}(b-c) have been adapted from Ref.~\cite{Su:2022p2358} with permission.(d) Half-order diffraction around ($H'$, $K'$, $L'$) and corresponding $\delta L = L-L'$ scans for a (LaPrNdSmEu)$_{0.2}$NiO$_3$ film on NdGaO$_3$, and (e) in-plane bond-angle distribution obtained from ABF-STEM imaging of the same film. The Fig.~\ref{Fig4}(d-e) have been adapted from Ref.~\cite{Bhattacharya:2025p2418490} with permission.}
\end{figure}

Under epitaxial strain, the complexity becomes even richer. Here, local chemical disorder-driven strain fields must contend with the long-range epitaxial strain enforced by the substrate, two structural directives acting simultaneously. This competition is clearly illustrated in (La$_{0.2}$Pr$_{0.2}$Nd$_{0.2}$Sm$_{0.2}$Eu$_{0.2}$)NiO$_3$ films grown under tensile and compressive strain by Bhattacharya \emph{et al.}~\cite{Bhattacharya:2025p2418490}. Under tensile strain exerted by NdGaO$_3$ substrates, half-order diffraction peaks confirm that the film inherits the substrate’s $a^-b^+c^-$ octahedral rotation pattern [Fig.~\ref{Fig4}(d)], demonstrating long-range crystalline uniformity. In contrast, ABF–STEM imaging shows a wide distribution of \textit{B}–O–\textit{B} angles [Fig.~\ref{Fig4}(e)], driven by $RE$-site compositional fluctuations, as corroborated by electron energy loss spectroscopy (EELS) mapping. These chemical fluctuations have been found to be intrinsically present irrespective of the choice of substrate. However,  the consequences of such nanoscale chemical inhomogeneity on electronic properties are highly  dependent on epitaxial strain, which are detailed exclusively in Section 6 of this review article. At a more extreme limit, chemical inhomogeneity can manifest as cation segregation. Wang \textit{et al.} showed that in Sr-doped $B$-site complex CCPO films, Sr substitution promotes Cr$^{3+}$ to Cr$^{6+}$ states, leading to the migration of the smaller, high-valence Cr ions during growth. EELS mapping revealed pronounced Cr segregation, likely to be driven by strain relief along with induced ionic radii mismatch, highlighting how epitaxial strain and multicomponent chemistry can cooperatively amplify compositional instabilities~\cite{Wang:2025p12719}.

While tolerance factor has classically served as the principal metric to predict perovskite stability, the increased chemical diversity in CCPOs brings forward additional determinants that guide how local coordination environments evolve. In this regard, Zhang \textit{\textit{et al.}}~\cite{Zhang:2023p2214} provided key insight by systematically exploring ${A}^5$TiO$_3$ perovskites [${A}^5$ indicates five members among Ba, Sr, Ti, Ca, Y, Na, Bi, K in equimolar fraction at $A$ site]. Their work revealed that the ability to stabilize a single-phase solid solution is highly sensitive to the ionic size disparity among the substituted cations. Notably, they also identified that electronegativity difference, rather than the tolerance factor, plays the dominant role in governing structural symmetry here: compositions with a difference below 0.4 preferentially adopt a cubic perovskite framework, whereas larger mismatches ($>$ 0.4) drive a transition toward a tetragonal distortion. This marks a crucial shift in structural design rules where cationic chemistry, rather than geometry alone, governs the accessible symmetry landscape in CCPOs.

The compositional diversity itself may create multiple locally preferred bonding environments, which drive the lattice to accommodate multiple structural symmetry clusters with different octahedral arrangements. This understanding directly connects to another important development, where such variations are used as a design strategy. In particular, local structural disorder, through octahedral distortions and polymorphic fluctuations, has emerged as a key structural basis for the superior dielectric behavior recently demonstrated in HEOs. Chen \textit{\textit{et al.}}~\cite{Chen:2022p3089} achieved a breakthrough in dielectric design by introducing a local polymorphic strategy in CCPOs by introducing numerous ions (Li$^{+}$, Ba$^{2+}$, Bi$^{3+}$, Sc$^{3+}$, Hf$^{4+}$, Zr$^{4+}$, Ta$^{5+}$, Sb$^{5+}$) with different ionic radii and valence states incorporated into K$_{0.2}$Na$_{0.8}$NbO$_3$ lattices. They demonstrate that rhombohedral (R), orthorhombic (O), tetragonal (T), and cubic (C) nanoclusters coexist here alongside randomly oriented octahedral tilts. This engineered mosaic of competing local symmetries leads to ultra-small polar nanoregions (PNRs), the implications will be discussed later in the context of dielectric properties. This polymorphic strategy has since been extended to BaTiO$_3$-based CCPOs, where coexisting T-R local environments are stabilized via compositional complexity~\cite{Zhang:2024p185}. In a related effort, Duan \textit{\textit{et al.}}~\cite{Duan:2024p6754} used high angle annular dark field imaging (HAADF–STEM) to show that entropy engineering in Bi$_{0.47}$Na$_{0.47}$Ba$_{0.06}$TiO$_3$ (BNBT)-based CCPOs. Atomic-resolution HAADF–STEM and selected area electron diffraction (SAED) analysis revealed entropy-driven coexistence of R, O, and T distortions, forming a dynamic ferroic landscape composed of fluctuating PNRs, featuring diverse \textit{B}O$_6$ tilt and polarization states, resulting in a superparaelectric-like response. 

Momentum space based EXAFS studies have also been performed for probing the local chemical environment in CCPOs. Sun \textit{\textit{et al.}}~\cite{Sun:2024p2406453} demonstrated in La(CoNiMgZnNaLi)$_{1/12}$Ru$_{1/2}$O$_3$ that despite random \textit{B'}-site occupation, all cations share a statistically identical octahedral environment with no site preference. Yet, the substantial ionic size mismatch induces heterogeneous local coordination distortions that promote charge redistribution from Ru toward Co/Ni via oxygen, strengthening orbital hybridization near the Fermi level. This study demonstrates that functional enhancement in CCPOs originates from locally distorted octahedra and the electronic consequences of disorder, rather than long-range crystallographic modification.

Overall, these investigations into local structure reveal that, despite exhibiting a globally single-phase crystal structure, CCPOs possess significantly more intricate local environments. This structural complexity at the atomic scale has profound implications for their physical properties and functional behavior, which are discussed in detail in subsequent sections of this review.

\section{Effect of compositional complexity on electronic properties}

\begin{figure}
    \centering	\includegraphics[width=.81\textwidth] {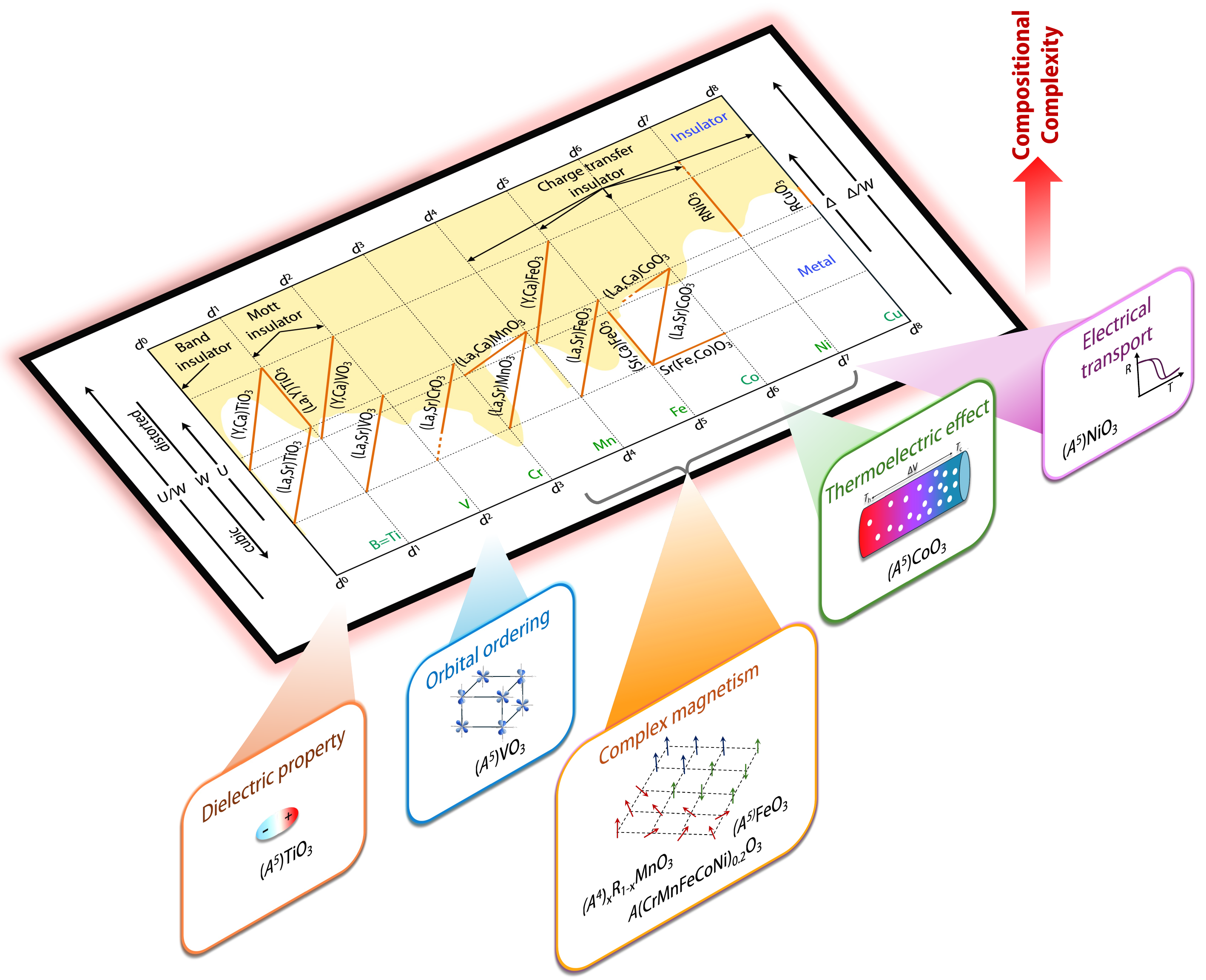}
	\caption{\label{Fig5} Electronic phase diagram for perovskite TMOs based on their correlated energy scales ($U$ and $\Delta$ scaled by the bandwidth $W$) and electronic configurations, adapted from Ref.~\cite{Imada:1998p1039}. The introduction of compositional complexity can be considered as the 3$^{rd}$ dimension to the phase space. Several families of CCPOs have recently been synthesized and investigated.}
\end{figure}

We now focus on the electronic properties of the CCPOs. The electronic structure of 3$d$ transition metal oxides (TMOs) differs markedly from that of conventional semiconductors. For example, LaTiO$_3$, with a $d^1$ electronic configuration, is predicted by band theory to be metallic; however, it is experimentally found to be an insulator due to strong electron correlation effects~\cite{Pavarini:2004p176403}. The electronic behavior of 3$d$ TMOs is often described using the Zaanen–Sawatzky–Allen (ZSA) phase diagram, which incorporates three key energy scales: on-site Coulomb repulsion ($U$), charge transfer energy ($\Delta$), and hopping interactions~\cite{Zaanen:1985p418}. The value of $U$ generally increases across the 3$d$ series from left to right in the periodic table, while $\Delta$ is strongly dependent on the oxidation state of the transition metal, typically decreasing with increasing oxidation state~\cite{Bocquet:1996p1161,Imada:1998p1039}. As a result of these competing energy scales, 3$d$ perovskite oxides can exhibit a variety of insulating phases, including Mott insulators, charge-transfer insulators, covalent insulators, and conventional band insulators [see Fig.~\ref{Fig5}]~\cite{khomskii:2014}. In addition, factors such as crystal field splitting, Hund’s coupling, and the strong interplay among spin, charge, lattice, and orbital degrees of freedom contribute to the emergence of a wide spectrum of electronic and magnetic phases. Indeed, nearly every known electronic or magnetic ground state has a representative within the perovskite oxide family. It is therefore intuitive that in CCPOs, local structural disorder and cationic heterogeneity, arising from $A$-site, $B$-site, or mixed sublattice complexity, would collectively reshape the underlying electronic energy landscape, leading to a large number of reports within a short span of time. All of these efforts of 3$d$ TM based CCPOs have been broadly categorized in Fig.~\ref{Fig5}. In the following sections, we discuss how compositional complexity influences fundamental physical properties (orbital ordering and electrical transport), as well as  its role in enabling enhanced functionalities such as improved dielectric response and reduced thermal conductivity. The influence of compositional complexity on magnetism has been discussed in details in the latter part of this paper.

\subsection{Electrical transport}
The interplay between disorder and electron correlation on electrical transport of quantum materials has been a subject of great interest for many decades~\cite{Anderson:1958p1492,Belitz:1994p261,Byczuk:2005p056404, Efros:2012book}. Therefore, it is only intuitive that CCPOs would provide an excellent platform to investigate and validate these longstanding ideas. Unfortunately, most of the reported HEOs exhibit exceedingly high or even immeasurable resistivity. This specially holds for the HEOs where the states near the Fermi level are contributed by the disordered sublattice. As a viable way to engineer metallic behavior within HEOs family, Patel \emph{\textit{et al.}} showed for the first time that in a CCPO-based $RE$NiO$_3$, complexity introduced at the $RE$ site can preserve a metallic phase at room temperature
~\cite{Patel:2020p071601}. Upon cooling, they observed that (La$_{0.2}$Pr$_{0.2}$Nd$_{0.2}$Sm$_{0.2}$Eu$_{0.2}$)NiO$_3$ [(LPNSE)NO] undergoes a first-order metal–insulator transition (MIT), with the transition temperature remaining close to that of the parent NdNiO$_3$, having a comparable tolerance factor. In this section, we build upon these findings to review the key studies on $RE$NiO$_3$-based CCPOs. We then further highlight a distinct class of $B$-site disordered CCPOs that display unconventional carrier density–transport characteristics while still maintaining measurable resistivity.

\textbf{Metal-insulator transition:} The $RE$NiO$_3$ series has long served as a canonical platform to study bandwidth-controlled MITs, where the electronic phase diagram is primarily governed by the $RE$ ionic radius through its effect on the tolerance factor and Ni–O–Ni bond angles, thereby modulating Ni–O hybridization and electronic bandwidth~\cite{Middey:2016p305}. In this series, only LaNiO$_3$ remains a paramagnetic metal down to very low temperatures, whereas the other members undergo an MIT that coincides with a bond-disproportionation transition. In NdNiO$_3$ (NNO) and PrNiO$_3$, the MIT is additionally accompanied by a magnetic transition from the paramagnetic state to an $E'$-type antiferromagnetic order (for details we refer to the reviews Ref.~\cite{Middey:2016p305,Chakhalian:2022p053004}).

While (LPNSE)NO, when grown under tensile strain (on NdGaO$_3$ substrate) shows a first-order MIT at $T_{\mathrm{MIT}} \sim 180$ K analogous to NNO film,  the electrical resistivity of the metallic phase was found to be higher compared to NNO ~\cite{Patel:2020p071601}, indicating the importance of local structural variations. EELS mapping along with bond angle maps directly displayed the presence of insulating patches within a metallic matrix [Fig.~\ref{Fig6}(a)], responsible for this lower electrical conductivity~\cite{Patel:2023p031407,Bhattacharya:2025p2418490}. Under compressive strain on SrLaAlO$_4$ (SLAO), however, this resistance enhancement of (LPNSE)NO over NNO is largely suppressed, and a robust metallic state persists down to low temperatures, similar to NNO on SLAO~\cite{Liu:2013p2714} [Fig.~\ref{Fig6}(b)]. These findings demonstrate that epitaxial strain can override disorder-driven scattering and restore electronic homogeneity. The suppression of the insulating phase under compressive strain was also investigated using hard X-ray photoemission spectroscopy (HAXPES), which revealed that an increase in Ni–O covalency is the prime factor for stabilizing this metallic state [Fig.~\ref{Fig6}(c),(d)]. Another study on the same composition under different strain conditions, emphasized the role of orbital polarization on the MIT~\cite{Cui:2023p115802}. Additionally,  introducing oxygen vacancies in this CCPO has emerged as an additional powerful tuning parameter, leading to a series of distinct electronic phases~\cite{Joshi:2025arXiv}.

\begin{figure}[b!]
    \centering	\includegraphics[width=0.95\textwidth] {Fig6}
         \caption{\label{Fig6}Electrical transport of CCPOs: (a) Schematic of coexisting insulating regions embedded within a metallic matrix in the overall metallic phase of (LPNSE)NO films. (b) Temperature-dependent sheet resistance of (LPNSE)NO grown on NGO (tensile strain) and SLAO (compressive strain).(c) Ni 2$p$ core-level HAXPES spectra for (LPNSE)NO on NGO and SLAO, where the separation between the main and satellite peaks (*) reflects on the Ni-O covalency. (d) Valence-band spectrum of (LPNSE)NO on SLAO, highlighting metallic character (0 marks the Fermi level); the shaded region corresponds to Ni $e_g^{*}$–O 2$p$ hybridized states. (e) Phase diagram of a high $RE$-cation variance (Y$_{0.2}$La$_{0.2}$Nd$_{0.2}$Sm$_{0.2}$Gd$_{0.2}$)NiO$_3$ film on LSAT substrate reproduced from ~\cite{Mazza:2023p013008} with permission. (f) Temperature-dependent resistivity plot of CCPO Sr$_{0.95}$(TiCrNbMoW)O$_3$ compared with SrVO$_3$, single-crystal SrMoO$_3$, and Sr$_{0.75}$NbO$_3$ adapted with permission from ~\cite{Almishal:2025p09868}.} 
\end{figure}

Apart from the tolerance factor, another crucial parameter governing the electronic properties of CCPOs is the cation size variance, defined as   $\sigma^2 = \sum_{i=1}^{n} y_i r_i^2 - \langle r_A \rangle^2$
where $r_i$ is the ionic radius of the $i$-th $RE^{3+}$ cation, $\langle r_A \rangle$ is the mean $A$-site ionic radius, and $y_i$ is its fractional occupancy~\cite{Rodriguez:1996p15622}. The variance quantifies the degree of local structural distortion arising from size mismatch, providing a measure of the lattice’s intrinsic configurational strain.  For (LPNSE)NO, $\sigma^2$ is $\sim$ 9.8 pm$^2$, whereas at the higher end of the variance spectrum lies (Y$_{0.2}$La$_{0.2}$Nd$_{0.2}$Sm$_{0.2}$Gd$_{0.2}$)NiO$_3$, exhibiting $\sigma^2 \sim 23.3$~pm$^2$. Mazza \textit{\textit{et al.}} investigated thin films of this composition grown on (LaAlO$_3$)$_{0.3}$(Sr$_2$TaAlO$_6$)$_{0.7}$ (LSAT) substrates displaying an intriguing behaviour beyond expectations, based solely on the average tolerance factor: while a $T_{\mathrm{MIT}}$ of $\sim$ 350 K is predicted, the observed transition occurs at $\sim$ 410 K~\cite{Mazza:2023p013008}. Moreover, an additional intermediate phase transition around 330 K, situated between the magnetic ordering and the MIT is reported, decoupled from both charge order and magnetism [Fig.~\ref{Fig6}(e)]. This anomalous behavior elucidates that increasing size variance not only modifies the bandwidth but can also stabilize emergent phases that surpass simple tolerance-factor considerations.

\textbf{Compositional complexity driven carrier crossover:}  In another report, Zhang \textit{et al.}~\cite{Zhang:2024p2406885} employed compositional complexity as a tunable design parameter in rare-earth cobaltates, La$_{1-x}$(Nd,Sm,Gd,Y)$_{x/4}$CoO$_3$, to investigate how atomic-scale disorder modulates spin-state transitions and carrier character in the absence of charge doping. Systematic variation of $A$-site cation mixing revealed that increasing complexity enhances crystallinity, suppresses defect density, and continuously tunes the Seebeck coefficient ($S$), resulting in an unexpected crossover from hole to electron-dominated transport. The induced lattice distortions within the CoO$_6$ framework preferentially localize valence-band holes over conduction-band electrons, shifting the Fermi level toward the conduction band and driving this intrinsic transition from $p$ to $n$ type behavior. This carrier-type reversal, achieved without aliovalent substitution, highlights an unconventional complexity-induced doping mechanism, offering a new paradigm for compositional complexity as an intrinsic control knob for electronic structure in correlated oxides.

\textbf{Transparent conductor with correlated electrons:} 
As we mentioned earlier in this article that the HEOs having disorder in TM sublattice are mostly insulating in nature. In this regard, a significant advancement has been reported very recently by Almishal \textit{\textit{et al.}}~\cite{Almishal:2025p09868}.
They investigated $B$-site disordered Sr$_{0.95}B^5$O$_3$ ($B^5$ = Ti$_{0.2}$Cr$_{0.2}$Nb$_{0.2}$Mo$_{0.2}$W$_{0.2}$) films grown on LSAT substrates  and demonstrated appreciable electrical conductivity. Moreover, those samples exhibited a markedly higher carrier concentration compared to their simpler perovskite counterparts — most of which are weakly metallic (except SrTiO$_3$, a band insulator).
The residual resistivity ratio (RRR) (ratio of the room temperature resistivity and low temperature resistivity), a key metric for assessing metallicity and disorder-driven scattering in complex oxides, is particularly informative in this case. In CCPO Sr$_{0.95}B^5$O$_3$ thin films, the RRR is lower compared to the parent perovskites [Fig.~\ref{Fig6}(f)]. This seemingly antagonistic coexistence of high carrier concentration and semiconducting transport suggests a highly crystalline lattice with strengthened electron correlations and chemical disorder, yielding local metallic regions separated by small transport barriers within a nonperiodic (or pseudoperiodic) disorder-induced potential landscape. Further optical measurements reveal a high visible transparency enabling optical band engineering through chemical disorder. This result is particularly interesting, because transparent materials are usually highly insulating while electrically conductive materials are typically opaque under the visible spectra~\cite{Khan:2020p1}, high-entropy design makes this rare combination possible.

Together, these studies demonstrate that compositional complexity provides a versatile means to tailor electronic transport in correlated oxides through the combined effects of chemical inhomogeneity, lattice distortion, disorder and electronic correlation. 
Moving forward, probing local-scale chemical inhomogeneity and its correlation with electronic phase behavior in high-variance CCPOs will be an interesting avenue to explore.

\subsection{Dielectric properties}

As previously discussed in Section 4, high-entropy design offers multiple routes to stabilize PNRs in dielectrics. The resulting effects on dielectric properties include broadened relaxation and enhanced coupling among electronic, ionic, and dipolar polarizations. Together, this enables high permittivity, low dielectric loss, and large breakdown strength in HEOs. A comprehensive overview of these mechanisms for HEOs has been presented by Chen \textit{et al.}~\cite{Chen:2023p6602}; here, we focus specifically on CCPOs, where these effects are particularly pronounced. As highlighted in the phase diagram in Fig.~\ref{Fig5}, the majority of dielectric studies carried out in HEOs fall under titanate-based compositions. One of the earliest works by Pu \emph{et al.}~\cite{Pu:2019p223901} reported relaxor-like behaviour in (Na$_{0.2}$Bi$_{0.2}$Ba$_{0.2}$Sr$_{0.2}$Ca$_{0.2}$)TiO$_3$. Subsequent studies on CCPO titanates, such as (Pb$_{0.25}$Ba$_{0.25}$Sr$_{0.25}$Ca$_{0.25}$)TiO$_3$, further revealed relaxor behavior with low dielectric loss, originating from small ferroelectric domain size, a reduced presence of 90$^\circ$ domain walls, and PNRs with short relaxation times~\cite{Xiong:2021p2979}. These findings emphasize the role of structural complexity in tuning relaxor responses within titanate CCPOs.

\begin{figure}[t!]
    \centering	\includegraphics[width=1.0\textwidth] {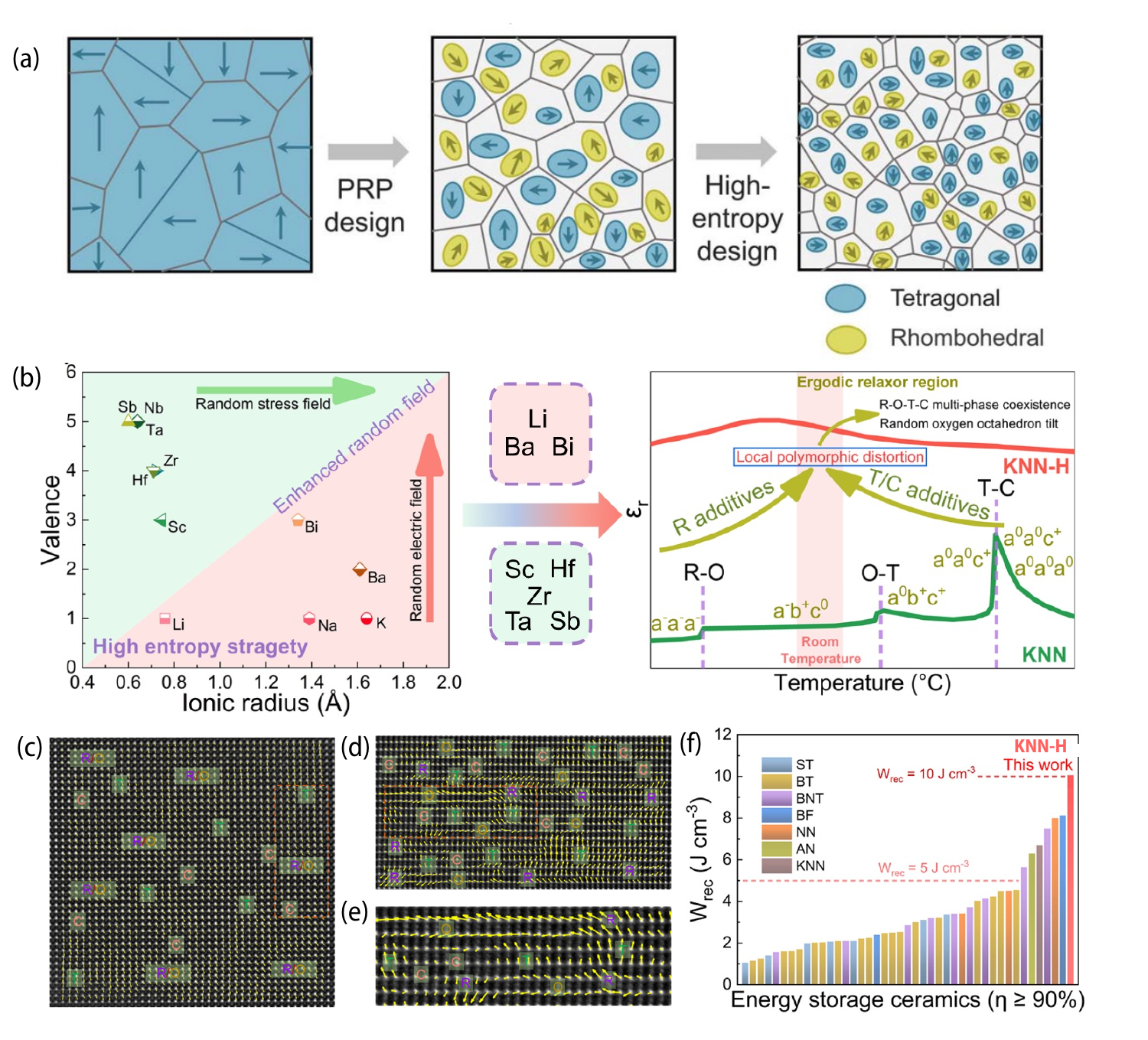}
	\caption{\label{Fig7} Dielectric design of CCPOs: (a) Schematic of the domain structure for the polymorphic relaxor phase (PRP) and the high-entropy design concept, adapted with permission from~\cite{Zhang:2024p185}. (b) Schematic illustration of the high-entropy design strategy for inducing local polymorphic distortions in the K$_{0.2}$Na$_{0.8}$NbO$_3$–based CCPO (KNN-H). (c) HAADF-STEM polarization vector map along [100]$_c$ and (d) [110]$_c$ for KNN-H. (e) Magnified view of the marked region showing the gradual transition of polarization vectors from T to R to O to C. (f) Comparison of energy storage density (W$_\mathrm{rec}$) (efficiency $\geq$ 90$\%$) for KNN-H ceramics with other reported lead-free bulk ceramics exhibiting high W$_\mathrm{rec}$. Panels (b)–(f) adapted with permission from~\cite{Chen:2022p3089}.}
\end{figure}

Chen \textit{et al.}~\cite{Chen:2022p3089} introduced another PNR engineering route in K$_{0.2}$Na$_{0.8}$NbO$_3$-based relaxors, where multiple aliovalent cations (Fig.~\ref{Fig7}(b),(c)) generate local polymorphic distortions (structural details in Section 4). As shown in Fig.~\ref{Fig7}(c)–(e), nanoscale structural clusters with gradually rotating polarization vectors (T $\rightarrow$ R $\rightarrow$ O $\rightarrow$ C) coexist within a C-phase matrix, disrupting long-range ferroelectric order and yielding smaller PNRs. These reduced-size PNRs lead to enhanced breakdown electric field and delayed polarization saturation. Consequently, an exceptional recoverable energy density with ultrahigh efficiency was achieved (Fig.~\ref{Fig7}(f)), establishing disorder-driven local symmetry fluctuations as a powerful strategy for optimizing dielectric performance.

A related advancement was reported by Duan \textit{et al.}~\cite{Duan:2024p6754} in high-entropy superparaelectrics (Local structural results have been discussed in section 4). Here, local polymorphic distortion resulted in diverse heterogeneous polarization states, leading to lowering of switching barriers, further promoting superparaelectric behavior. This approach simultaneously enables high polarization response, minimal remnant polarization, delayed saturation, and enhanced breakdown electric fields, further demonstrating the versatility of the polymorphic distortion strategy.

Building upon this, the polymorphic distortion-based design was logically extended to multilayer ceramic capacitors (MLCCs), which are highly sought after in modern electronics~\cite{Covaci:2022p3869}. In a BaTiO$_3$-based CCPO MLCC, multiple $A$-site dopants (Na$^+$, Ca$^{2+}$, Sr$^{2+}$, Sm$^{3+}$) and $B$-site Zr$^{4+}$ stabilize a polymorphic relaxor phase by embedding rhombohedral BiFeO$_3$ into tetragonal BaTiO$_3$~\cite{Zhang:2024p185}. This configuration reduces hysteretic loss, lowers switching barriers, and improves dielectric breakdown strength—greatly enhancing MLCC efficiency and operational reliability.

In another widely used class of lead-free ferroelectric energy-storage materials, Bi(Mg$_{0.5}$Ti$_{0.5}$)O$_3$ (BMT)~\cite{Xie:2019p3819}— known for its strong ferroelectricity and structural stability—compositional complexity has also been introduced to further tune dielectric properties. The CCPO system applied a similar PNR-focused strategy as discussed in Section 4, exhibiting a relaxor-like response and achieving nearly an eightfold enhancement in energy density, along with improved insulation, reduced leakage current, and higher breakdown strength~\cite{Li:2024p4940}. Similar enhanced energy-storage characteristics have been reported in other Bi-based titanates with $A$-site complexity~\cite{Yang:2022p1083,Wei:2025p807,Bunpang:2024p6017}, further validating the effectiveness of complexity-driven relaxor engineering. Interestingly, in the CCPO (Bi$_{0.2}$Na$_{0.2}$K$_{0.2}$La$_{0.2}$Sr$_{0.2}$)TiO$_3$, Wei \emph{et al.} ~\cite{Wei:2025p807} emphasized the crucial role of short-range order in enhancing energy-storage performance. Their study also showed that apart from coexisting C-R-T phases, Nb doping at the Ti site further induces short-range chemical ordering, which significantly improved the energy-storage density and efficiency. 

 A natural question that follows is how the degree of compositional complexity influences dielectric behavior—specifically, whether increasing entropy always enhances performance. To address this, Xiong \textit{et al.}~\cite{Xiong:2024p2415351} investigated La$B$O$_3$ ($B$ site cation have been selected among Fe, Co, Ni, Cu, Mn) by systematically varying the number and combination of $B$-site cations to tune configurational entropy. They observed a non-monotonic, parabolic dependence of dielectric relaxation, with the strongest response at medium entropy in La(Fe$_{0.33}$Co$_{0.33}$Ni$_{0.33}$)O$_3$ (LFCNO). Low entropy limits lattice distortion, whereas excessive disorder at high entropy suppresses long-range dipole alignment, thereby weakening cooperative polarization. Additionally, the optimized disorder at intermediate entropy promotes stronger field-induced polarization and dielectric attenuation, enabling efficient electromagnetic absorption with thin structures and wide operating ranges. Similar medium-entropy behavior is observed in $A$-site complex $A^5$TiO$_3$, where vacancy-driven dipoles lead to a large complex permittivity, suppressing wave penetration~\cite{Liu:2024p2400059}. Thus, medium-entropy CCPOs offer an emerging route to electromagnetic shielding—an uncommon functionality in conventional dielectric systems.

Overall, these studies establish compositional complexity as a powerful lever to engineer local structural distortions, suppress long-range ferroelectricity, and create highly dynamic PNRs—unlocking exceptional dielectric energy-storage behavior that is difficult to achieve in conventional materials. Interestingly, however, the very mechanisms that enable these enhancements introduce a delicate balance: too little disorder yields weak polarization response, while excessive disorder frustrates dipole cooperativity. Thus, optimizing the degree of compositional complexity is a key to realizing robust and reliable high-performance dielectrics. Furthermore, the PNR design strategy is also being actively investigated in the light of enhancing piezoelectric performance in CCPOs~\cite{Liu:2022p118115, Wang:2025p22166}.

\subsection{Thermoelectric properties}
We next move to the performance of CCPO-based thermoelectric materials, as thermoelectricity is an important topic for green energy harvesting~\cite{He:2017peaak9997}. 
The conversion efficiency of a thermoelectric device is given by its figure of merit $zT = \frac{S^2\sigma T}{\kappa}$, where $S$ is the Seebeck coefficient, $\sigma$ is the electrical conductivity and $\kappa$ is the combined electronic and lattice thermal conductivity. Perovskite oxides such as SrTiO$_3$, CaMnO$_3$, etc. have shown to be promising thermoelectric materials due to their high-temperature stability, abundance, and low processing cost~\cite{Jana:2025p27050}. However, their high lattice-dominated thermal conductivity (typically 5 - 10 W\,m$^{-1}$K$^{-1}$) limits their thermoelectric efficiency. 
Strategies like doping, composite formation, and grain refinement have effectively reduced thermal conductivity but not below the chalcogenide benchmark. Since approaches like nanostructuring are less effective in oxides with inherently short phonon mean free paths, alternative design strategies are essential~\cite{Jana:2025p27050}.
One such strategy involves leveraging compositional complexity, as extreme chemical disorder can induce strong phonon scattering, often bringing lattice thermal conductivity close to the amorphous limit~\cite{Hu2:2018p1802116, Jiang:2022p208,Tang:2024p1641}.

The exploration of thermoelectricity in HEOs began with a CCPO itself that incorporated complexity at the $B$-site, Sr(Ti$_{0.2}$Fe$_{0.2}$Mo$_{0.2}$Nb$_{0.2}$Cr$_{0.2}$)O$_3$~\cite{Banerjee:2020p17022}.
In this work, Banerjee \textit{\textit{et al.}} demonstrated that ultralow thermal conductivity arises from enhanced phonon scattering induced by multiple transition-metal occupancies at the $B$-site. The resulting local disorder introduces point defects and Anderson localization, giving rise to semiconductor-like transport governed by a small-polaron hopping mechanism. Consequently, both $S$ and $\sigma$ increase with  $T$, within the range of their measurement.

In titanate-based HEOs, compositional complexity offers an elegant framework to decouple carrier and phonon transport through complexity-driven local structural distortions. These distortions generate a disordered potential landscape that strongly scatters phonons, thereby suppressing lattice thermal conductivity, while maintaining relatively coherent electronic transport via locally optimized carrier pathways. For instance, thin films of (Sr$_{0.2}$Ba$_{0.2}$Ca$_{0.2}$Pb$_{0.2}$La$_{0.2}$)TiO$_3$ exhibit a dramatic reduction in lattice thermal conductivity which approaches the amorphous limit, while controlled Ti off-centering enhances carrier mobility~\cite{Zheng:2024p7650}. This entropic modulation of local structure effectively decouples charge and heat transport, leading to a significantly improved thermoelectric response and an enhanced $zT$. Entropy engineering in CCPO manganites have also shown to enhance thermoelectric performance through synergistic optimization by utilizing configurational disorder to strengthen phonon scattering (reducing $\kappa$) and simultaneously improving electron transport (increasing $\sigma$)~\cite{Shi:2022p15582}. 
Zhang \emph{et al.}~\cite{Zhang:2023p168366}, demonstrated  another thermoelectric design strategy with (Sr$_{0.25}$Ca$_{0.25}$Ba$_{0.25}$$RE_{0.25}$)TiO$_3$ ($RE$= Nd, Sm, Eu, Gd, Dy, Ho) wherein a temperature-independent, low glass-like $\kappa$ was realized over a wide temperature range, embodying a ‘phonon-glass electron-crystal’ concept, arising from the combined influence of a large TiO$_6$ octahedral distortion and a complex local strain field. The $zT$ value was also found to be dependent on the choice of the $RE$. Another work by Lou \emph{et al.}~\cite{Lou:2022p3480} demonstrated that although a combination of octahedral tilt, antiparallel cationic displacement, oxygen vacancies and edge dislocations reduce the $\kappa$ significantly, the Anderson localization effect significantly reduces $\sigma$ as well, thus, decreasing the overall $zT$ value. Thus, although the introduction of compositional complexity offers a promising paradigm for designing thermoelectrics, the understanding the local structure and defect chemistry is essential to achieve high $zT$ value, suitable for industrial deployment~\cite{Tang:2024p1641,Zhou:2021p1378,Jiang:2022p208,Bhui:2025p29542}.

\subsection{Orbital ordering}
Orbital ordering (OO) and its impact on charge, spin and lattice is a well known concept in modern condensed matter physics, that has garnered significant attention in the family of TMOs~\cite{Tokura:2000p462,Khomskii:2021p2992, Khomskii:2022p054004}. The rare-earth vanadates ($RE$VO$_3$) have long served as a canonical platform for understanding OO~\cite{Miyasaka:2003p100406,Zhang:2025p026508}. Accordingly, compositionally complex $RE$VO$_3$ became a natural starting point to explore orbital phenomena in CCPOs~\cite{Yan:2024p024404}. Systematic studies across different $RE$ combinations reveal that both the average ionic radius and its variance critically dictate the nature and sequence of orbital and magnetic transitions, much like what we discussed for the case of electrical transport. Compositions with small ionic variance and intermediate ionic radii (like: $RE$ = La, Ce...,Tb) preserve the conventional $G$-type orbital and $C$-type antiferromagnetic order, akin to their parent compounds. Whereas, larger size variance suppresses or merges these transitions, leading to the emergence of spin–orbital entangled states. Notably, systems with comparable mean radii but differing variance display distinct transition pathways from multi-step orbital and magnetic ordering to a single, entangled phase transition, underscoring the decisive role of configurational fluctuations. Thus, in these $RE$VO$_3$-based CCPOs, compositional complexity does not merely average structural effects but actively reshapes the orbital landscape, offering a new route to engineer correlated states within the Kugel–Khomskii framework. Although the study of OO in CCPOs is still at an early stage, it would be particularly interesting to investigate CCPOs with active $e_g$ orbitals.

\begin{figure}[t!]
    \centering
	\includegraphics[width=0.95\textwidth] {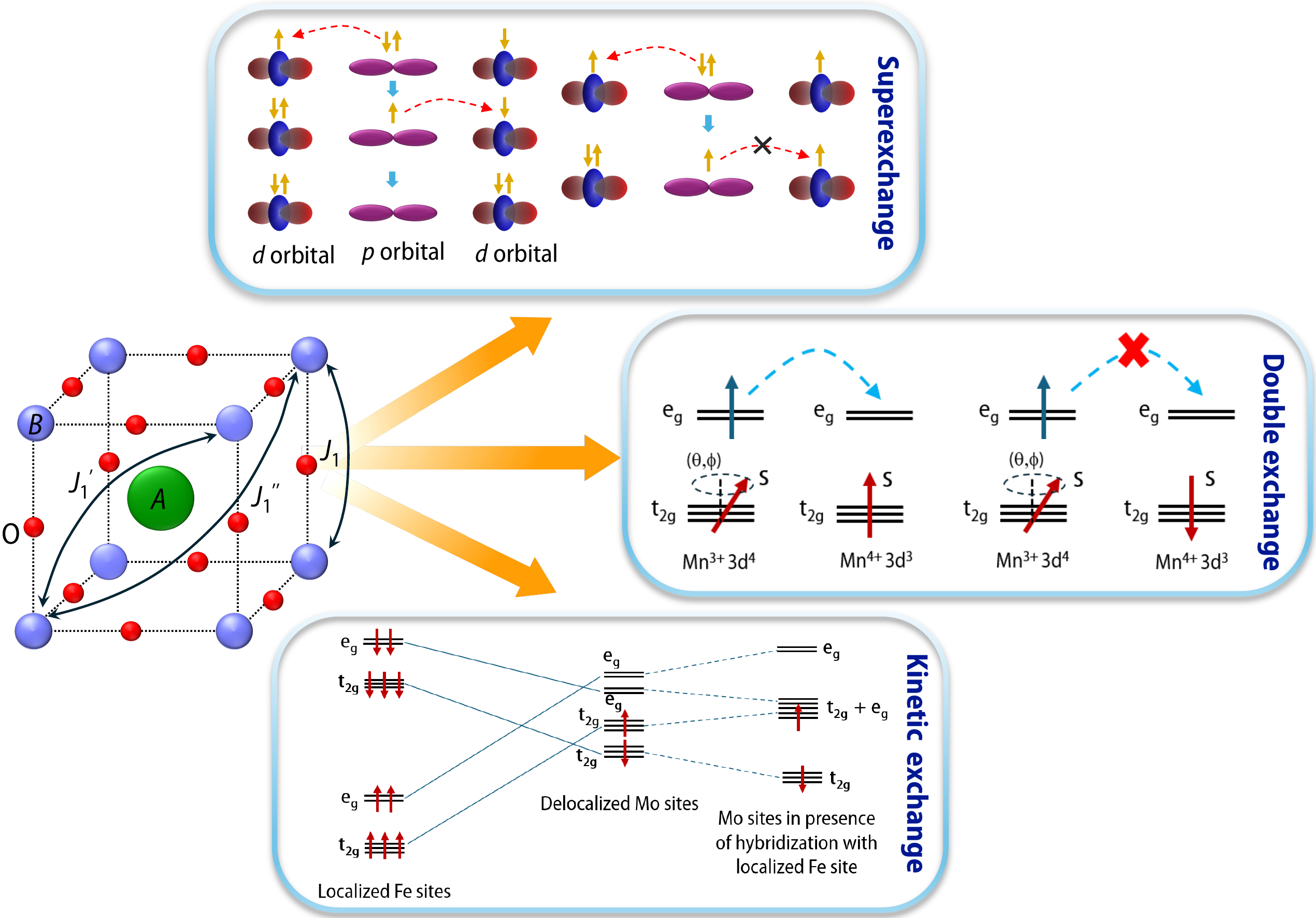}
         \caption{\label{Fig8} Perovskite crystal highlighting the different neighbouring magnetic interactions along with schematic representation of superexchange, double exchange and kinetic energy driven interactions. The model of kinetic exchange for Sr$_2$FeMoO$_6$ has been adapted from Ref.~\cite{Sarma:2000p2549}. The exchange coupling for Mo$^{5+}$ is enhanced in presence Fe-Mo hopping interactions. For details, we refer to Ref.~\cite{Sarma:2000p2549}.} 
\end{figure}

\section{Effect of compositional complexity on magnetism}
Magnetism has been known to humankind for over a thousand years, yet its scientific understanding has evolved primarily during the past century~\cite{Coey:2010book}. Early investigations focused on TMOs, particularly binary oxides~\cite{Roth:1958p1333}. Over time, perovskite oxides emerged as a central class of materials for developing fundamental concepts, owing to the rich diversity and complexity of magnetic interactions in 3$d$ transition-metal-based perovskite systems~\cite{Zener:1951p403,Goodenough:1955p564,khomskii:2014}. Consequently, the concept of introducing compositional complexity within the perovskite lattice has garnered significant attention~\cite{Witte:2019p034406}, resulting in a rapid surge of publications in recent years. While earlier works up to 2021 were comprehensively reviewed by Sarkar \emph{et al.}~\cite{Sarkar:2021p1973}, this article emphasizes recent developments while revisiting some key earlier studies for continuity and completeness.

In perovskite structures, direct exchange between two $B$-site ions is absent; all magnetic interactions are indirect exchange, mediated through intermediate oxygen atoms [Fig.~\ref{Fig8}]. The first-neighbor $B$-O-$B$ interaction ($J_1$) occurs along the side linkage, while the second-neighbor interaction ($J_1'$) involves two $B$ ions located diagonally across a face of the unit cell via two oxygen atoms. The third-neighbor interaction ($J_1''$) connects two $B$ ions, located along the body diagonal [See Fig.~\ref{Fig8} for interactions]. Among all indirect exchange mechanisms, superexchange is the most common. It's nature [ferromagnetic (FM) or antiferromagnetic (AFM)] and strength depend on the bond angle and the electronic configurations of $B$ and $B'$ ions, as described by the Goodenough–Kanamori–Anderson (GKA) rules~\cite{KanamoriL:1959p87,Anderson:1950p350,Roth:1958p1333}. For instance, $180^\circ$ TM–O–TM bonds between half-filled orbitals favor AFM coupling [Fig.~\ref{Fig8}], whereas overlaps involving half-filled and empty or filled orbitals, often near $90^\circ$ geometries, can promote FM alignment. Goodenough extensively tabulated these interactions for transition-metal ion pairs in his seminal book `Magnetism and the Chemical Bonds'~\cite{Goodenough:1963book}.
Another important mechanism is the double exchange (DE), which becomes relevant in mixed-valence systems~\cite{Anderson:1955p675}. For example, ferromagnetic DE between Mn$^{3+}$ and Mn$^{4+}$ in hole-doped manganites is stronger than superexchange, resulting in a ferromagnetic metallic phase [Fig.~\ref{Fig8}]~\cite{Tokura:2006p797}. In recent years, strong evidence has emerged for an additional mechanism, kinetic energy driven magnetic exchange in many TMOs [Fig.~\ref{Fig8}]~\cite{Sarma:2000p2549,Saha-Dasgupta:2024p023001}. Introducing compositional complexity is therefore expected to disrupt, modify, or introduce new magnetic exchanges, often leading to spin frustration. Consequently, phenomena such as spin-glass behavior, inhomogeneous magnetism, and exchange bias, etc. would naturally arise. Any observation beyond these expectations, even the appearance of a long-range magnetic order is noteworthy and warrants careful consideration.

\begin{figure}[!b]
    \centering	\includegraphics[width=.95\textwidth] {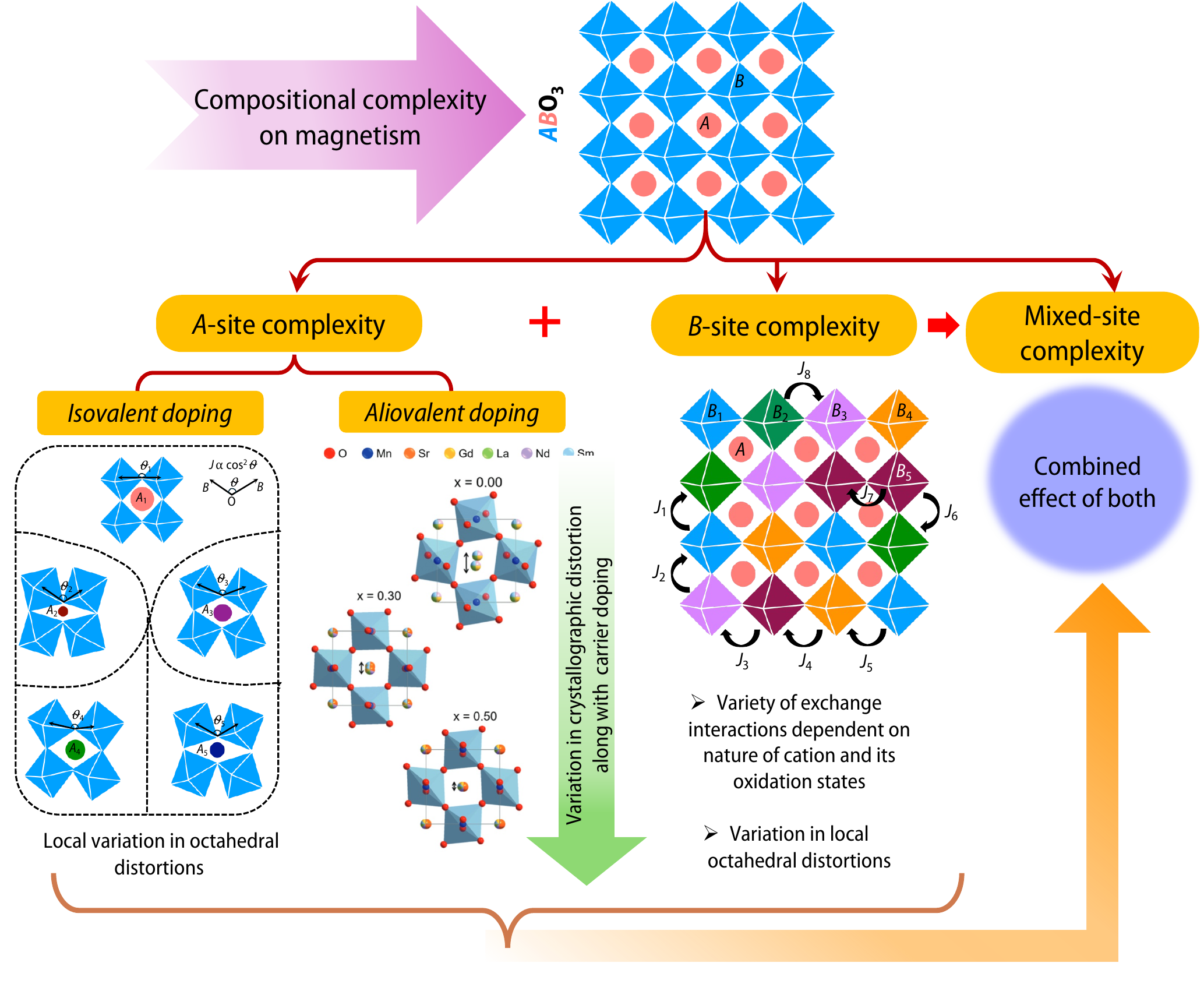}
         \caption{\label{Fig9}  Microscopic influence of compositional complexity within the perovskite lattice on magnetic properties.  The figure under aliovalent doping has been adapted with permission from Ref.~\cite{Sarkar:2023p2207436}.} 
\end{figure}

With these basic background, we focus on the works on magnetism in CCPOs which can be broadly categorized into three classes as shown in Fig.~\ref{Fig9}. Firstly, the compositional complexity at the $A$ site, where isovalent substitution induces ionic-size mismatch and local strain, promoting octahedral tilts and bond-angle deviations [Fig.~\ref{Fig9}], thereby indirectly tuning the magnetic exchange strength without altering $B$-site electronic configuration. In contrast, aliovalent $A$-site substitution introduces both size disorder and charge imbalance, creating mixed-valence states at the $B$-site and directly modifying TM–O–TM exchange~\cite{Sarkar:2023p2207436}. Secondly, the compositional complexity at $B$-site further introduces multiple, nonequivalent exchange pathways ($J_1$, $J_2$, …) between cations of different valence and orbital occupancy [Fig.~\ref{Fig9}]. When disorder is present at both $A$ and $B$ sites, coupled charge disorder, lattice distortions, and competing interactions generate a highly heterogeneous magnetic energy landscape with spatially varying octahedral geometries and exchange strengths.

\subsection{Compositional complexity at \textit{A} site.}

The role of $A$-site compositional complexity in CCPO systems has been systematically evaluated across $B$-site  cations ranging from Cr to Ni. Owing to the distinct magnetic characteristics exhibited by each $A^5B$O$_3$ class, the discussion is organized according to increasing 3$d$ electron count. Given the extensive body of work on manganites, this subgroup is discussed in a dedicated subsection.

\textbf{\emph{RE}$^5$\emph{B}O$_3$ ($B$ = Cr, Fe, Co, Ni):} 
Cr-based CCPO ($RE^5$CrO$3$ with $RE^5$=La$_{0.2}$Nd$_{0.2}$Sm$_{0.2}$Gd$_{0.2}$Y$_{0.2}$) systems exhibit canted antiferromagnetic order, analogous to their parent counterparts~\cite{Witte:2020p185109}. At low temperatures, a spin-reorientation transition (SRT) induces magnetization reversal via antiferromagnetic $RE$–Cr coupling, as observed in GdCrO$_3$ but absent in other orthochromites (e.g., Sm, Nd). This similarity has been attributed to the largest moment of the Gd$^{3+}$ and its strong coupling to the Cr sublattice. Additionally, Li \emph{et al.}~\cite{Wenyong:2023p1413} reported exchange-bias effects in the $RE^5B$O$_3$ with $RE^5$= La$_{0.2}$Pr$_{0.2}$Nd$_{0.2}$Sm$_{0.2}$Y$_{0.2}$, underscoring the need for further microscopic investigations.

Progressing to $\mathrm{Fe}$-based CCPO, orthoferrites ($RE$FeO$_3$) are well known for their canted $G$-type AFM ordering, $RE$-dependent SRTs, and low-temperature $RE$-moment ordering~\cite{Tokunaga:2008p097205,Lee:2011p117201,Chen:2012p103905,Bousquet:2016p123001}. In this context, the ($A^5$)FeO$_3$ CCPO analogues retain the parent-like robust magnetic ground state, indicating that $A$-site disorder does not significantly weaken Fe$^{3+}$-O$^{2-}$ orbital overlap. This resilience originates from dominant $\sigma$-type superexchange in the $3d^5$ configuration, which is less sensitive to local distortions than $\pi$-type interactions~\cite{Witte:2020p185109}. Nevertheless, the strongly enhanced coercive fields—comparable to highly distorted LuFeO$_3$, highlight the role of local distortion fields induced by compositional complexity~\cite{Witte:2020p185109}. Moreover, by altering the specific $A$-site constituents ($A^5$ = Tm$_{0.2}$Nd$_{0.2}$Dy$_{0.2}$Y$_{0.2}$Yb$_{0.2}$), the tunable SRT range increases by an order of magnitude relative to parent orthoferrites, offering enhanced control over magnetic anisotropy and magnetocaloric performance~\cite{Yang:2024p23203}. AC susceptibility studies also reveal spin-glass behavior in $RE^5$FeO$_3$ ($RE$ = Gd, Dy, Ho, Er, Tb), originating from competing FM and AFM interactions between the disordered $RE$ and Fe sublattices~\cite{Yin:2022p082404}, similar to other compositionally complex oxides~\cite{Pramanik:2024p3382}.

Co-based CCPO systems inherit the intrinsic complexity of parent $RE$CoO$_3$ compounds, arising from multiple spin states of Co$^{3+}$ ($d^6$): low spin (LS, $S$ = 0), intermediate spin (IS, $S$ = 1), and high spin (HS, $S$ = 2). This variability often leads to contrasting results across synthesis routes and studies. Witte \emph{et al.} reported a predominantly LS state in $RE^5$CoO$_3$~\cite{Witte:2020p185109}, whereas Krawczyk \emph{et al.} observed a gradual LS $\rightarrow$ IS $\rightarrow$ HS crossover~\cite{Krawczyk:2020p3211}. Future investigations on high-quality single crystals and epitaxial thin films could provide critical insights into disorder-driven spin-state transitions.

In high-variance nickelates such as (Y$_{0.2}$La$_{0.2}$Nd$_{0.2}$Sm$_{0.2}$Gd$_{0.2}$)NiO$_3$, whose electrical properties were discussed in Section 6.1, the system undergoes a paramagnetic-to-$E^\prime$-type antiferromagnetic transition at a temperature consistent with predictions based on the parent phase diagram and average tolerance factor~\cite{Mazza:2023p013008}.

Overall, the magnetic behavior of these CCPOs with $A$-site complexity largely parallels that of their parent compounds, although local distortions introduce subtle variations.
Expanding the $A$-site complexity to double perovskite family, a recent study by Bhattacharya \textit{et al.} demonstrated that in a high-variance $RE_2$NiMnO$_6$ compound, a strong FM ground state is still preserved, with $T_\mathrm{c}$ dictated primarily by $\langle r_A \rangle$ rather than $\sigma^2$ and a mean field approach serves as a good starting point~\cite{Bhattacharya:2025p02485}.

\textbf{\emph{RE}MnO$_3$ based CCPOs:} The $RE$MnO$_3$ family hosts a remarkably rich magnetic phase diagram governed by Mn–O–Mn superexchange, where tuning the $RE$ cation drives transitions among $A$-type AFM, $E$-type AFM, and spiral (spin-cycloidal) states~\cite{Kimura:2003p060403}, establishing these systems as model platforms for compositional-complexity–driven magnetism. The N\'eel temperature ($T_\mathrm{N}$) systematically scales with the $A$-site ionic radius: it is highest for La due to a larger Mn–O–Mn bond angle, and decreases for smaller ions such as Eu that induce stronger octahedral distortions. Building on this, Kumar \textit{et al.} explored 
(La$_{0.2}$Nd$_{0.2}$Pr$_{0.2}$Sm$_{0.2}$Eu$_{0.2}$)$_{1-x}$Sr$_x$MnO$_3$ as a function of $x$~\cite{Kumar:2023p101026}.
Their detailed measurements also point to the role of average ionic radii of the $A$-site on the MIT and magnetic transition temperature [Fig.~\ref{Fig10}(a)]. However, the sample with $x$=0.5 showed multiple transition.

The effect of compositional complexity in 50\% hole-doped manganite was further systematically examined by Das \emph{et al.} considering Nd$_{0.5}$Sr$_{0.5}$MnO$_3$~\cite{Das:2023p169950}. Introducing multiple trivalent cations at the $A$ site in (La$_{0.1}$Pr$_{0.1}$Nd$_{0.1}$Gd$_{0.1}$Bi$_{0.1}$)Sr$_{0.5}$MnO$_3$ increases average ionic radii ($\langle r_A \rangle$) and reduces variance ($\sigma^2$), yet counterintuitively suppresses FM-DE even under structurally favorable conditions. Further variation in the trivalent cation set leads to either a complete loss of FM order with a stabilized charge-ordered AFM state, or, when both the fraction of Nd and Sr sites are varied keeping 50\% hole doping intact, a fully suppressed long-range order resulting in an AFM ground state without charge order. These results demonstrate that magnetism is strongly dictated by local lattice distortions and cation disorder, beyond simple $\langle r_A \rangle$–$\sigma^2$ correlations. 

In a subsequent (La$_{0.6}$Sr$_{0.4}$)MnO$_3$-based CCPO~\cite{Das:2023p024411}, simultaneous trivalent/divalent substitution allowed tuning of both $\langle r_A \rangle$ and $\sigma^2$ while keeping the Mn$^{3+}$/Mn$^{4+}$ ratio fixed. Here, the conventional rise of $T_\mathrm{c}$ with increasing $\langle r_A \rangle$ collapses, revealing $\sigma^2$-driven magnetic control. Notably, strong FM ordering persists without AFM or glassy phases, attributed to local distortions that disrupt DE pathways into FM clusters while maintaining ferromagnetism globally. Supporting this trend, further studies on $RE^5$MnO$_3$ systems with smaller $RE$ cations such as Eu, Ho, and Yb showed that enhanced octahedral distortions (reduced $\langle r_A \rangle$ and tolerance factor) correlate with stronger magnetic saturation, reaffirming the central role of structural distortion~\cite{Qin:2024p0272}. 

Significant advances in magnetic functionalities such as CMR and exchange bias have also been realized in $RE$MnO$_3$-based CCPO systems. Sarkar \textit{et al.} demonstrated that Sr-induced hole doping in (Gd$_{0.25}$La$_{0.25}$Nd$_{0.25}$Sm$_{0.25}$)$_{1-x}$Sr$_x$MnO$_3$ enables a controlled evolution of magneto-electronic phases—from insulating canted-AFM at low doping, through phase-separated CMR states, to a robust metallic DE-FM state at higher doping. The exceptionally high CMR values already surpass those of conventional bulk manganites, highlighting the strong tunability offered by high-entropy design. In another study, Nathan \textit{et al.} showed that in (Y$_{0.2}$La$_{0.2}$Pr$_{0.2}$Nd$_{0.2}$Tb$_{0.2}$)MnO$_3$, the low-temperature regime exhibits intrinsic domain pinning and measurable exchange bias under high magnetic fields~\cite{Nathan:2024p13474}, suggesting strong potential for further enhancement through avenues such as doping or strain engineering.

\subsection{Compositional complexity at \textit{B} site.}
 The introduction of compositional complexity at the $B$-site adds an additional hierarchy of exchange pathways and competing energy scales [Fig.~\ref{Fig9}]. In a CCPO containing five different cations on the $B$-site sublattice, each $B$-element can form over 200 distinct nearest-neighbor combinations~\cite{Mazza:2022p2200391}, raising a fundamental question: can long-range magnetic order emerge under such conditions?  
The first investigation in this direction was conducted by Witte et al. in 2019, who explored a series of compounds with the general formula
$RE$[Co$_{0.2}$Cr$_{0.2}$Fe$_{0.2}$Mn$_{0.2}$Ni$_{0.2}$]O$_3$ with $RE$= La, Gd, Nd, Sm, and Y were investigated~\cite{Witte:2019p034406}. 
We primarily focus on La based compound, as this composition has been extensively studied in subsequent works. In bulk form, LaCrO$_3$, LaMnO$_3$, LaFeO$_3$ are orthorhombic and undergoes antiferromagnetic transitions with a $T_\mathrm{N}$ of 290 K, 100 K and 740 K. In contrast, the other two members LaCoO$_3$ and LaNiO$_3$  are rhombohedral and paramagnetic. Bulk La[Co$_{0.2}$Cr$_{0.2}$Fe$_{0.2}$Mn$_{0.2}$Ni$_{0.2}$]O$_3$ [abbreviated as La$B^5$O$_3$], synthesized by the nebulized spray pyrolysis method, adopts the orthorhombic structure.  The observation of a non-saturating hysteresis loop with a large coercive field, along with the presence of exchange bias, points to magnetic inhomogeneity— most likely arising from ferromagnetic clusters embedded within an AFM matrix. Complementary STEM+EELS measurements on the same samples revealed chemical fluctuations (also discussed in Section 4)~\cite{Su:2022p2358}, which may be responsible for the observed inhomogeneous magnetic behavior.

Parallel efforts by researchers at Oak Ridge National Laboratory have focused on stabilizing the same composition in single-crystalline form and investigating its structural and magnetic properties through a series of studies~\cite{Brahlek:2020p054407,Sharma:2020p014404,Mazza:2021p094204,Mazza:2022p2200391}. Bulk samples were synthesized via solid-state reaction, while thin films were grown using pulsed laser deposition. Detailed X-ray diffraction analyses confirmed that the thin films were highly homogeneous and adopted an orthorhombic phase~\cite{Brahlek:2020p054407}. Notably, the resulting structure was distinct from both the individual parent compounds and their averaged structure. STEM+EELS mapping further revealed a uniform distribution of $B$-site cations, indicating atomic-scale disorder that is spatially homogeneous~\cite{Sharma:2020p014404}. Neutron diffraction measurements identified the emergence of a G-type antiferromagnetic phase in both bulk and thin film samples~\cite{Mazza:2022p2200391}, with transition temperature consistent with that observed in samples prepared via spray pyrolysis~\cite{Witte:2019p034406}. Crucially, the single-crystalline thin films exhibited no magnetic relaxation. This emergence of magnetic order from a disordered configuration is striking, prompting a deeper philosophical question: in such a disordered system, does each spin experience its own unique local interaction, or does it respond only to an effective mean field? Recent studies on various CCOs lend strong support to the applicability of the mean-field approach in describing their magnetic behavior~\cite{Nevgi:2025p7038,Sharma:2024p224432,Bhattacharya:2025p02485}.

\begin{figure}[t!]
    \centering
	\includegraphics[width=1.0\textwidth]{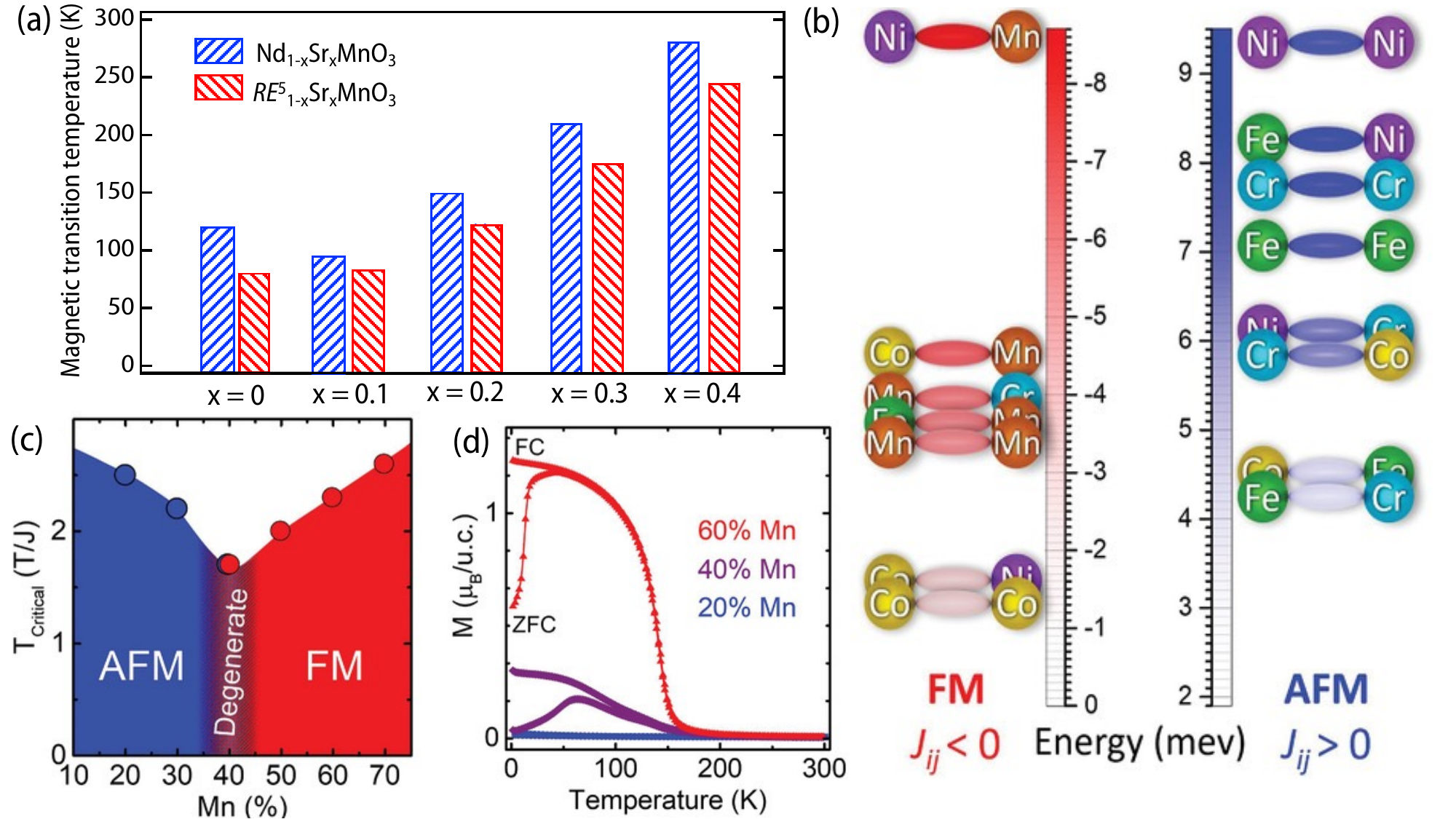}
         \caption{\label{Fig10} Magnetic properties of CCPOs: (a) Comparative magnetic transition temperatures of (La$_{0.2}$Nd$_{0.2}$Pr$_{0.2}$Sm$_{0.2}$Eu$_{0.2}$)$_{1-x}$Sr$_x$MnO$_3$ and Nd$_{1-x}$Sr$_x$MnO$_3$, which share similar tolerance factors, for doping levels $x = 0$, 0.1, 0.2, 0.3, and 0.4. (b) Magnetic ordering tendencies and associated energy scales for La-based TMOs, illustrating phase-space modulation through combinations of $B$-site cations. (c) Monte Carlo–derived phase diagram for La(Cr$_y$Mn$_x$Fe$_y$Co$_y$Ni$_y$)O$_3$ ($y = (1 - x)/4$), shown as a function of Mn content. (d) Field-cooled (FC) and zero-field-cooled (ZFC) magnetization versus temperature under a 1 kOe magnetic field for synthesized films. Panel (a) created using data from~\cite{Kumar:2023p101026}, and panels (b)–(d) adapted with permission from~\cite{Mazza:2022p2200391}.} 
\end{figure}

To gain microscopic insight into the long-range magnetic order, Mazza \emph{et al.}~\cite{Mazza:2022p2200391} also performed Monte Carlo simulations using a classical Heisenberg model. The simulations revealed a percolated AFM state, consistent with the long-range AFM order observed in experiment. The dominant AFM interactions—primarily involving Fe with Fe, Cr, Co, and Ni. In contrast,  Mn favors ferromagnetic coupling [Fig.~\ref{Fig10}(b)]. Despite FM bonds comprising roughly 40\% of the total interactions, the system exhibits a single transition temperature. The authors argued that regions with ferromagnetic interactions coexist within the AFM matrix but remain incoherent and non-percolating, appearing at the same critical temperature without forming a long-range FM phase. Together, these experimental and simulation results suggest that critical temperature and the nature of magnetic ordering  can be broadly engineered by considering  the average of the spin ($S$) and exchange parameters (($J$)) of constituent elements. Since Mn-O-$B$ connections prefer FM, further compositions were examined by varying Mn concentration while keeping the other four $B$ site ions in equiatomic fraction. Simulations predict that at 40\% Mn concentration both FM and AFM are degenerates while for Mn concentration larger than 50\%, unpercolated AFM clusters form within fully percolated FM matrix [Fig.~\ref{Fig10}(c)].  Magnetic measurements on single-crystalline films with 40\% and 60\% Mn confirmed these predictions [Fig.~\ref{Fig10}(d)].  Further investigation into the charge states of $B$-site cations as a function of Mn concentration would be crucial, especially in light of XAS measurements on La$B^5$O$_3$ by Wang \emph{et al.}~\cite{Wang:2023p58643}, which identified Cr$^{3+}$, Co$^{3+}$, Fe$^{3+}$, Ni$^{2+}$, and Mn$^{4+}$. Follow-up studies on Tb$B^5$O$_3$ and Dy$B^5$O$_3$ thin films revealed the presence of Co$^{2+}$, suggesting charge redistribution among $B$-site cations~\cite{Farhan:2022pL060404,Cocconcelli:2024p134422}, while a separate study on Lu$B^5$O$_3$ reported a significantly lower Co$^{2+}$ concentration~\cite{Farhan:2023p044402}.

Recognizing Co's pivotal role, Regmi \emph{et al.}~\cite{Regmi:2024p025023} substituted Ni and Cr with non-magnetic Ti and V to study its influence. In Sr(Fe$_{0.2}$Mn$_{0.2}$Co$_{0.2}$Ti$_{0.2}$V$_{0.2}$)O$_3$ thin films, a ferrimagnetic ground state emerged from antiparallel alignment of Co$^{2+}$ moments with those of Fe and Mn. These results highlight Co$^{2+}$ as a key driver of magnetic behavior and demonstrate that targeted substitution can effectively tune magnetic properties of CCPOs.

\subsection{Compositional complexity at both $A$ and $B$ site.}

Building on the discussion of magnetic properties arising from $A$- and $B$-sublattice complexity, a natural question is, what occurs when both sites exhibit compositional disorder. Witte \emph{et al.}~\cite{Witte:2019p034406} addressed this by studying $A^5B^5$O$_3$ systems ($A^5$ = (LaNdGdSmY)$_{0.2}$, $B^5$ = (CrMnFeCoNi)$_{0.2}$). Remarkably, despite extreme cationic disorder, these materials retain long-range AFM order. Moreover, $T_\mathrm{N}$ still correlates systematically with $A$-site ionic radius (tolerance factor), underscoring that even at maximal configurational entropy, lattice geometry governs dominant magnetic coupling. The broad magnetic transition observed was attributed to local distortions and a wide distribution of TM–O–TM bond angles~\cite{Su:2022p2358}. 
While this vast compositional phase space offers exciting avenues to design and derive exotic magnetism, it also makes a bruteforce approach towards such exploration impractical. Thus, establishing guiding principles is essential for identifying which compositions are  promising to pursue.

\section{Outlook}

In this review, we have highlighted how the physical properties of CCPOs emerge from a delicate interplay between local disorder and global structural parameters. While disorder-driven phenomena dominate many responses at the local scale, it is remarkable that several long-range macroscopic behaviors persist despite the highly disordered backdrop. Yet, identifying reliable structural descriptors, such as tolerance factor, variance, or their combinations, that predict material responses remains an unresolved challenge. To address this, a synergistic approach combining combinatorial synthesis, high-throughput mapping of multicomponent spaces, artificial intelligence and data-driven modeling— strategies that have transformed the discovery of HEAs~\cite{Murty:2019y2019,Wang:2025p100993}, will be equally critical for CCPOs. For instance, recent studies employing computational descriptors such as enthalpy of mixing ($\Delta H_{\mathrm{mix}}$) and bond-length distribution ($\sigma_{\mathrm{bonds}}$) have successfully captured the stability of all rocksalt-structured HEOs~\cite{Sivak:2025p216101}. These advances underscore that realizing the full potential of CCPOs requires a tightly coupled integration of theory and experiment through iterative feedback loops. Extending such data-driven design principles to CCPOs could unlock new pathways for controlling correlated electronic and magnetic ground states, ultimately leading to phases unique to the high-entropy landscape.

Another key challenge for CCPOs moving forward is the controlled engineering of electronic conductivity.
Recent efforts indicate that targeted band-gap engineering may offer a viable path forward: in rocksalt HEOs, for example, anion doping within a cation-disordered lattice has been shown to tune band gaps and thereby modulate electrical conductivity~\cite{Siniard:2025p05789}. Beyond compositional strategies, electrostatic or ionic gating may enable percolative conduction pathways, creating spatially heterogeneous resistance states with potential for memristive applications~\cite{Tsai:2023p2302979}. Furthermore, recent advances in thin-film growth position the field to exploit geometrical lattice engineering, an approach that has already demonstrated remarkable success in conventional TMOs for stabilizing exotic phases such as quantum spin liquids and topological states, polar metal, etc~\cite{Liu:2016p133,Middey:2016p056801,Kim:2016p68,Chakhalian:2020p050904,Liu:2021p2010}. 
Harnessing these directions will be pivotal for unlocking novel functionalities in CCPOs.

Other exciting avenue lies in the realization of exotic frustrated magnetic states, such as spin-liquid-like and spin-ice, like regimes—through compositional engineering~\cite{Zhou:2017p025003,Bramwell:2020p374010}. By leveraging the intrinsic statistical distribution of exchange couplings, CCPOs may offer a fundamentally distinct route compared to traditional geometrical frustration~\cite{Mustonen:2018p1085}. Furthermore, combining “exchange-variance engineering” enabled by compositional complexity with geometrical frustration could open an entirely new landscape of magnetic phenomena.

 A major unexplored area is the time-domain response of CCPOs and, more broadly, HEOs which can be probed using advanced pump–probe techniques~\cite{Hughes:2021p353001,Zhang:2014p19,Zong:2023p224,Murakami:2025p035001,Kirilyuk:2010p2731}. Perovskite oxides often exhibit multiple simultaneous transitions, making it challenging to identify the primary driving mechanism under steady-state conditions. In such scenarios, time-resolved pump–probe measurements become indispensable for disentangling coupled dynamics and revealing the underlying processes.
 For example, in NdNiO$_3$, femtosecond excitations revealed distinct recovery timescales for magnetic and electronic orders, enabling identification of the primary driver of the phase transition~\cite{Stoica:2022p165104}. Extending such approaches to CCPOs would be transformative. One particularly promising avenue is the direct tracking of phonon damping under these severe lattice distortion landscapes, where these probes may offer a unique pathway to map how phonon lifetimes, electron dynamics, and local structural relaxations unfold in femtosecond to nanosecond scales. Extending these techniques to magnetic CCPOs would provide insights into the role of spin-phonon coupling within the compositional complexity settings.
 
Overall, the rapid progress of CCO research over the last ten years is truly remarkable, establishing it as a major standalone theme within condensed matter physics, materials science, and chemistry. We believe that many exciting discoveries remain to be uncovered, achievable through a close synergy between experiment, theory, modeling, simulation and data-driven approaches.

\section*{Acknowledgements}
We thank Venkatraman Gopalan, A. R. Mazza and Le Wang for insightful comments.
NB and SD acknowledge funding from the Prime Minister’s Research
Fellowship (PMRF), MoE, Government of India. RN thanks the Indian Institute of Science for support through Sir C. V. Raman postdoctoral fellowship program.  SM acknowledges funding
support from a SERB Core Research grant (Grant No. CRG/2022/001906)
and I.R.H.P.A Grant No. IPA/2020/000034.

\section*{Disclosure statement}
The authors declare no competing interests.


\end{document}